\documentclass[useAMS,usenatbib]{mn2e}
\usepackage{graphicx,psfig,cite,epsf}  
\usepackage{times}
\usepackage{subfigure}
\usepackage{textcomp}
\usepackage{multirow}
\usepackage{amsmath}

\def\apjl{ApJ}
\def\apjl{ApJ}                
\def\apjs{ApJS}               
\def\ao{Appl.~Opt.}           
\def\aap{A\&A}                

\title[NGC 5643 ULX1: a BH of stellar origin?]{The ultraluminous X-ray source NGC 5643 ULX1: a large stellar mass black hole accreting at super-Eddington rates?}
\author[Fabio Pintore]{Fabio Pintore$^{1}$\thanks{pintore@iasf-milano.inaf.it,\newline fabio.pintore@dsf.unica.it},  Luca Zampieri$^2$, Andrew D. Sutton$^3$, Timothy P. Roberts$^4$, \newauthor Matthew J. Middleton$^5$, Jeanette C. Gladstone$^6$ \\
$^1$ INAF-Istituto di Astrofisica Spaziale e Fisica Cosmica - Milano, via E. Bassini 15, I-20133 Milano, Italy \\
$^2$ INAF-Osservatorio Astronomico di Padova, Vicolo dell'Osservatorio 5, I-35122 Padova, Italy \\
$^3$ Astrophysics Office, NASA Marshall Space Flight Center, ZP12, Huntsville, AL 35812, USA \\
$^4$ Department of Physics, University of Durham, Durham DH1 3LE, UK \\
$^5$ Institute of Astronomy, Madingley Road, Cambridge CB3 OHA, UK \\
$^6$ Department of Physics, University of Alberta, 11322-89 Avenue, Edmonton, AB T6G 2G7, Canada
}
\begin{document}

\maketitle

\begin{abstract}

A sub-set of the brightest ultraluminous X-ray sources (ULXs), with X-ray luminosities well above $10^{40}$ erg s$^{-1}$, typically have energy spectra which can be well described as hard power-laws, and short-term variability in excess of $\sim10\%$. This combination of properties suggests that these ULXs may be some of the best candidates to host intermediate mass black holes (IMBHs), which would be accreting at sub-Eddington rates in the {\it hard} state seen in Galactic X-ray binaries. 
In this work, we present a temporal and spectral analysis of all of the available \textit{XMM-Newton} data from one such ULX, the previously poorly studied 2XMM J143242.1$-$440939, located in NGC 5643. We report that its high quality EPIC spectra can be better described by a broad, thermal component, such as an advection dominated disc or an optically thick Comptonising corona. 
In addition, we find a hint of a marginal change in the short-term variability which does not appear to be clearly related to the source unabsorbed luminosity. We discuss the implications of these results, excluding the possibility that the source may be host an IMBH in a {\it low} state, and favouring an interpretation in terms of super-Eddington accretion on to a black hole of stellar origin. 
The properties of NGC 5643 ULX1 allow us to associate this source to the population of the {\it hard/ultraluminous} ULX class.

\end{abstract}


\section{Introduction}

Ultraluminous X-ray Sources (ULXs) are extragalactic, point-like, non-nuclear sources that are characterised by isotropic X-ray luminosities greater than $10^{39}$ erg s$^{-1}$ (e.g. \citealt{fabbiano89}), which is close to the Eddington limit of a 10 M$_{\odot}$ black hole (BH).
Observational evidence strongly suggests that the majority of ULXs are accreting BH X-ray binaries (XRBs) with massive donors (e.g. \citealt{zampieri09,fengsoria11}; and references therein). Luminosities higher than $\sim$3$\times10^{39}$ erg s$^{-1}$ rule out Eddington limited accretion onto typical stellar-mass BHs (sMBHs, $4 - 20$ M$_{\odot}$), and the non-nuclear locations of ULXs in their host galaxies rules out accretion on to supermassive BHs. 
Despite this, it has been proposed that ULXs may be powered by accretion on to either sMBHs (\citealt{king01}) or massive stellar BHs formed in low-metallicity environments ($\leq 80$ M$_{\odot}$; \citealt{zampieri09,mapelli09,belczynski10}). Both of these scenarios would require that many ULXs must be accreting matter at or above the Eddington limit \citep[e.g.][]{ohsuga11}, possibly with a degree of beaming further boosting their radiative emission (e.g. \citealt{king01, begelman06, poutanen07, king09}). 
In at least two ULXs it has been possible to constrain the mass of the compact object to within the stellar-mass and massive stellar regimes (5--30 M$_{\odot}$ in \citealt{liu13} and $\sim$14 M$_{\odot}$ in \citealt{motch14}), whilst in one other case the first robust evidence of a neutron star in a ULX has been reported \citep[M82 X-2,][]{bachetti14}.

Amongst ULXs, those brighter than $10^{41}$ erg s$^{-1}$ (e.g. HLX-1 in ESO 243-49; \citealt{farrell09}) are difficult to explain within this framework and are instead good candidates to host IMBHs ($10^2-10^5$ M$_{\odot}$; e.g. \citealt{colbert99}). These would be accreting in the canonical sub-Eddington regimes, which produce the observed states in Galactic XRBs \citep[i.e. {\it hard} and {\it soft} states; e.g.][]{mcclintock06, fender12}. However, potential formation and feeding mechanisms of IMBHs are still a matter of debate. It has been suggested that they may originate from either the collapse of primordial Population III stars \citep{madau01}; the dynamical collapse of supermassive stars forming in the centre of very dense stellar super-clusters \citep{portegies04}; grow from sMBHs that undergo repeated mergers in a dense globular cluster \citep{miller02}; or via a merger with a dwarf galaxy that expelled a nuclear IMBH \citep{farrell12, mapelli12}. 
In all these scenarios ULX activity can only start after the capture of a massive star, able to provide the required mass transfer to power the extreme ULX luminosities (e.g. \citealt{hopman04, mapelli11}).
Hence, identifying IMBHs is of crucial importance not only for simply proving their existence and assessing these formation scenarios, but also due to the likely implications for our understanding of the initial seeding and subsequent growth of nuclear BHs in galaxies \citep{volonteri10}.

\begin{figure}
\center
\includegraphics[width=5.5cm,height=5.5cm]{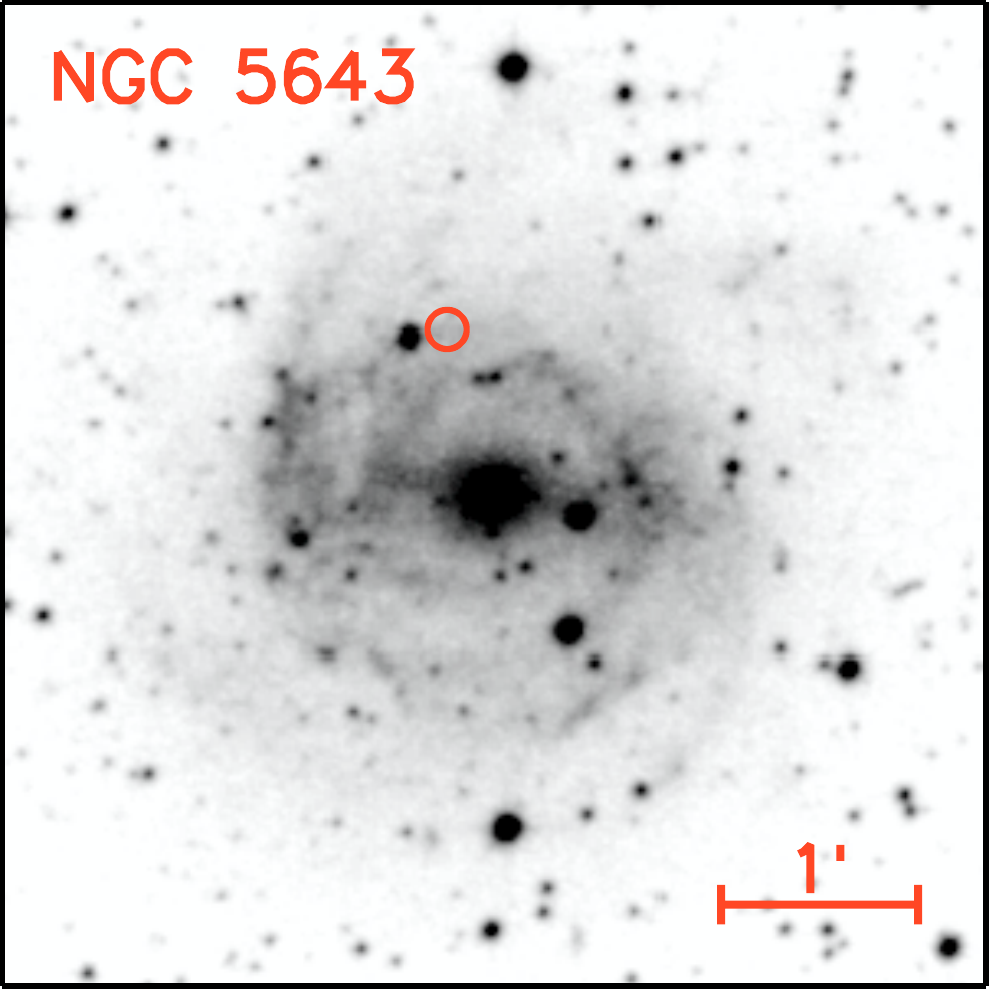}
\caption{DSS2 Blue image of the host galaxy of the
target ULX, with the position of the ULX indicated by a red circle (north is up). Note that the circle size is not indicative of the far smaller position error on the object.}
\label{opt_im}
\end{figure}

High quality \textit{XMM-Newton} X-ray spectra of ULXs with luminosities lower than $\sim 2\times10^{40}$ erg s$^{-1}$ show ubiquitous spectral curvature at moderately high energy ($> 2$ keV, \citealt{stobbart06, gladstone09}), often coupled to a soft excess of $\sim 0.1-0.4$ keV.
As these features are not commonly observed in Galactic XRBs, they imply a new spectral regime, which is usually referred to as the ``ultraluminous state'' \citep{roberts07, gladstone09}, and may be associated with super-Eddington accretion. A few luminous ULXs have very recently been observed with \textit{NuSTAR}, resulting in spectra which demonstrate that the high energy curvature persists above 10 keV \citep{bachetti13, walton13,rana14}, confirming this result. 
The spectral curvature may arise either from Comptonisation in a cold, optically thick corona lying above an accretion disc (e.g. \citealt{stobbart06,gladstone09}); from a modified accretion disc, i.e. a slim disc \citep{mizuno07}; or the hot inner regions of the disc, which are highly distorted by advection, turbulence, self-heating and spin \citep{beloborodov98,suleimanov02, kawaguchi03}. The soft component is thought to originate from the photosphere of a radiatively-driven wind that is  ejected from the spherization radius of the accretion disc down to its innermost regions, which is expected to occur at super-Eddington accretion rates (e.g. \citealt{shakura73,poutanen07,ohsuga07, ohsuga09}). Partially-inhomogeneous winds in ULXs may explain the  puzzling short-term variability reported in some ULXs \citep{heil09}, through a combination of inclination angle and accretion rate (e.g. \citealt{middleton11,takeuchi13,middleton15}). These same factors also determine the observed energy spectrum \citep[e.g.][]{sutton13,pintore14}. 

On the other hand, HLX-1 in  ESO 243-49 \citep{farrell09}, which has a peak luminosity of $\sim 10^{42}$ erg s$^{-1}$, is observed with a strikingly different set of properties than standard ULXs. 
It displays a spectral evolutionary pattern similar to that of the Galactic BH X-ray transients (e.g. \citealt{mcclintock06, fender12}), undergoing periodic outbursts \citep{godet14,webb14} in which the source switches from a high luminosity \textit{soft} state to a fainter \textit{hard} state \citep{servillat11}. 
Multi-wavelength emission has been detected from HLX-1 as it undergoes this out-burst cycle. In particular, radio emission has been detected, which is associated with discrete jet ejection events \citep{webb12}, similar to those that have been observed during {\it hard}-to-{\it soft} state transitions state in Galactic BHs. By assuming that HLX-1 was in a similar state, the authors of that study were able to refine the BH mass estimates to $\sim10^3-10^4$ M$_{\odot}$.
Other good IMBH candidates include: M82 X-1 \citep{feng10}, for which a BH mass of $\sim400$ M$_{\odot}$ has been suggested based on the tentative identification of twin-peak quasi periodic oscillations \citep{pasham14};
and an X-ray source in the galaxy NGC 2276 \citep{mezcua15}. Also, a sub-sample of moderately-high luminosity ULXs ($L_{X}$$>$5$\times10^{40}$ erg s$^{-1}$) show hard power-law like spectra, with no high energy turn-off and short-term variability at the level of $\sim10\%$ \citep{sutton12}. These sources may possibly be IMBHs in a \textit{hard} state, although their apparently flat spectral shape may simply be the result of poor counting statistics. 
\begin{figure*}
\center
\subfigure{\includegraphics[width=5.5cm,angle=-90]{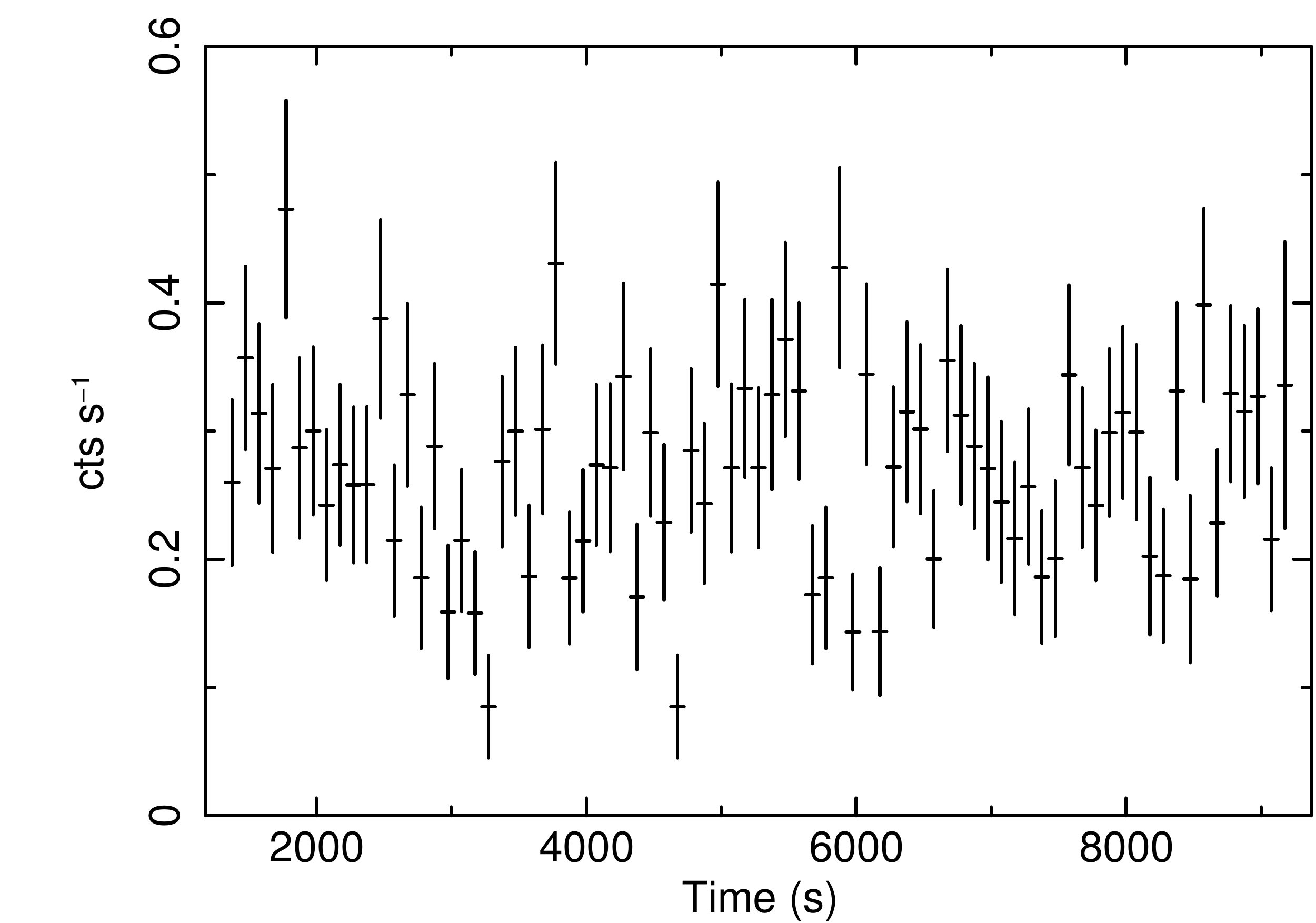}}
\subfigure{\includegraphics[width=5.5cm,angle=-90]{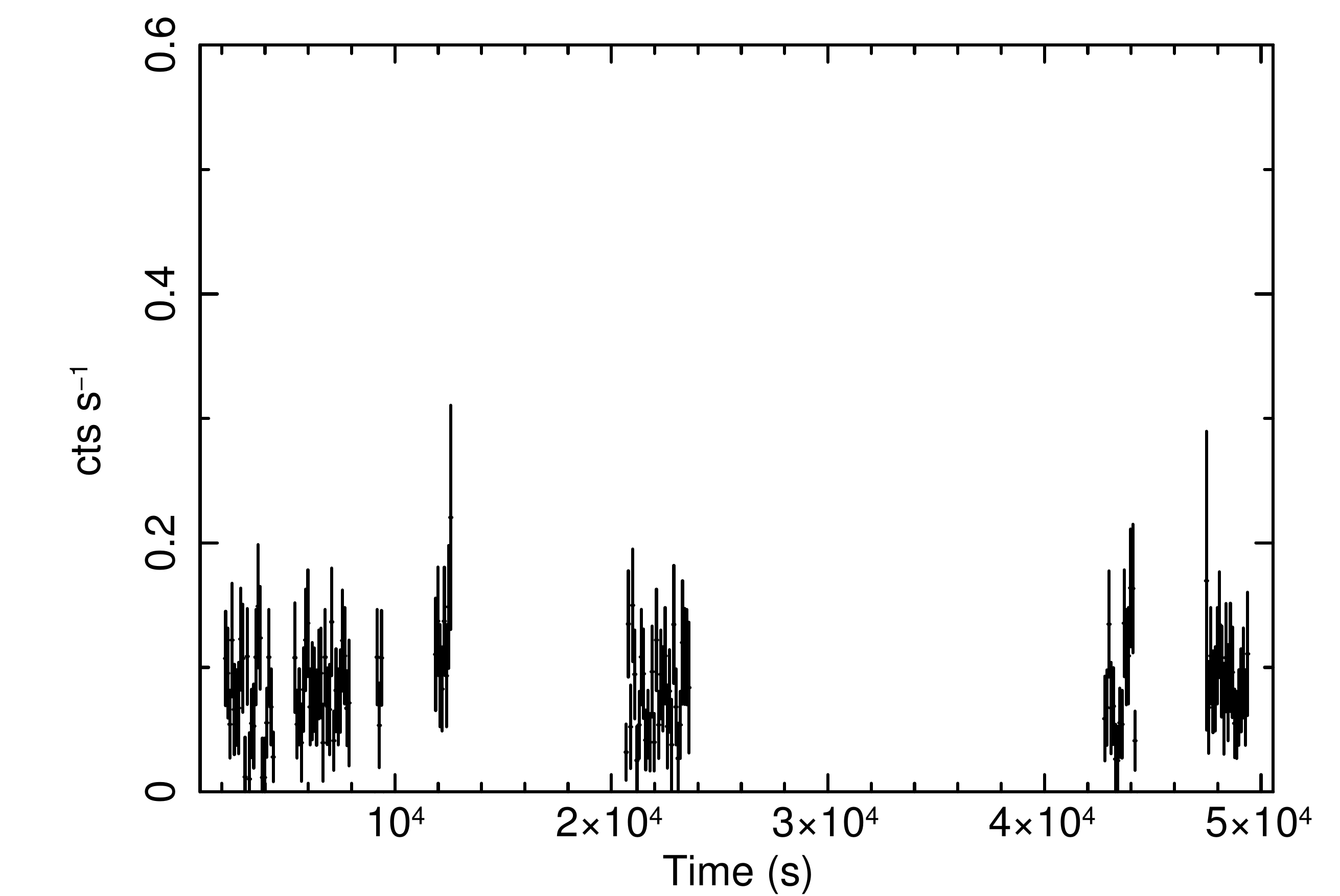}}
\subfigure{\includegraphics[width=5.5cm,angle=-90]{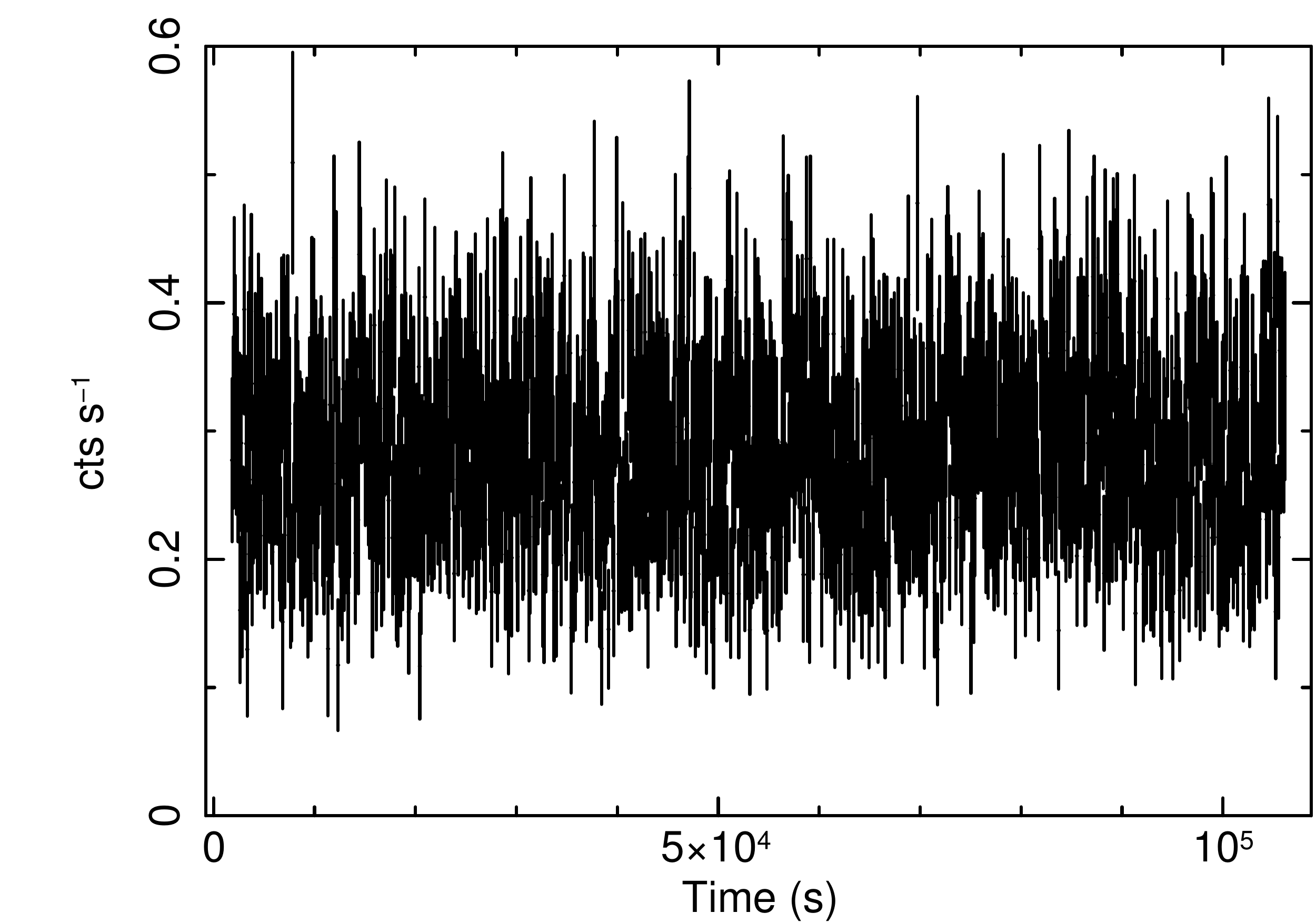}}
\caption{EPIC-pn 0.3--10 keV background subtracted lightcurve of NGC 5643 ULX1 extracted from observations 1 (\textit{top-left}) and 3 (\textit{bottom}), and an EPIC-MOS 0.3--10 keV lightcurve (obtained by stacking MOS1 and MOS2 data) from observation 2 (\textit{top-right}), all of which are sampled with time bins of 100 s. Gaps in the lightcurves are due to the removal of high energy particle flares.}
\label{lc}
\end{figure*}

\citet{walton11} carried out a cross-correlation between the 2XMM-DR3 and the RC3 catalogues, revealing a number of new extreme ULXs ($L_X>10^{40}$ erg s$^{-1}$). The target of this study (2XMM J143242.1$-$440939) is the nearest source from a preliminary extension of that work (Earnshaw et al. in prep.). 2XMM J143242.1$-$440939 is located in NGC 5643, which is an SAB(rs)c galaxy with a low-luminosity Seyfert II nucleus and strong star formation in its spiral arms \citep{phillips83}. NGC 5643 is at a distance of $d = 16.9$ Mpc \citep{tully88}, and 2XMM J143242.1-440939 (hereafter, NGC 5643 ULX1) is located 0.8 arcminutes from its nucleus (\citealt{guainazzi04}; Figure~\ref{opt_im}). The observed peak X-ray luminosity of NGC 5643 ULX1 is $\geq 10^{40}$ erg s$^{-1}$, although its historical flux record shows it has varied in luminosity by a factor of $\sim$3--4 (see the following sections). Recent X-ray observations with low counting statistics were consistent with it having a hard power-law continuum spectrum ($\Gamma \sim 1.7$, \citealt{guainazzi04,matt13}) and short-term fractional variability at the level of $\sim 10\%$. In this work we perform a spectral and temporal analysis of all {\it XMM-Newton} observations of the source, including our new very long observation taken on August 28 2014. We make use of the high quality data from this {\it XMM-Newton} observation ($\sim100$ ks) to test  whether we can rule out either sub- or super-Eddington accretion in NGC 5643 ULX1. 
Besides the ULX, NGC 5643 hosts also an AGN \citep{guainazzi04,bianchi06,annuar15}. Results of the spectral analysis of the AGN are also briefly reported in and discussed in Appendix~\ref{AGN}.

The paper is structured as follows: in Section~\ref{data_reduction} we summarize the data selection and reduction procedures, in Section~\ref{spectral_fits} we present the results of our timing and spectral analysis of NGC 5643 ULX1, and in Section~\ref{discussion} we discuss the implications of these.

\begin{table}
  \begin{center}
\footnotesize
   \caption{Log of the \textit{XMM-Newton} observations of NGC 5643 ULX1.} 
      \label{tab1}
   \begin{tabular}{l c c c c c }
\hline 
No. &  Obs.ID.  & Date &Exp. & Net Exp. $^a$ & Counts\\
 &   & &(ks)& (ks)& \\
\hline
1 &0140950101  & 2003/02/08 &9.7& 9.4  &1480$^b$\\
2 &0601420101  & 2009/07/26 &54.5& 12.0  & 690$^c$\\
3 &0744050101  & 2014/08/28 &116.9& 104.0  & 20466$^b$\\
\hline
\end{tabular}
\end{center}
$^a$ Net exposure time after removing high energy particle contamination; 
$^b$ total number of EPIC-pn counts; $^c$ total number of EPIC-MOS counts.
\end{table}

\section{Data Reduction}
\label{data_reduction}

NGC 5643 ULX1 was observed once with both the \textit{ROSAT} (August 28 1997) and \textit{Chandra} (December 26 2004) satellites, and three times with \textit{XMM-Newton} (see log in Table~\ref{tab1}). In this work we mainly focus on the \textit{XMM-Newton} observations, but also briefly consider the \textit{Chandra} data. The \textit{ROSAT} dataset was largely excluded, as it was taken with the HRI detector, which has negligible energy resolution. 

The \textit{Chandra} observation was reduced using the {\sc ciao} v.4.6 software adopting the standard procedures. In order to verify the point-like nature of NGC 5643 ULX1, we used {\sc chart} and {\sc marx} data analysis software to simulate the PSF for the observation at the source position. We used ten times the real exposure time to better sample the PSF. The {\sc ciao} tool {\it srcextent} was used to confirm that the source is not extended at 90$\%$ confidence. 
However, we found that the \textit{Chandra} data were affected by $\sim6\%$ pile-up (estimated with the {\sc ciao} v.4.6 software) and the data quality and spectral properties of NGC 5643 ULX1 were comparable to that of the \textit{XMM-Newton} observation 2, thus do not provide significant additional information. Therefore, in the rest of this work we focus mainly on the \textit{XMM-Newton} data. The observed \textit{ROSAT} and \textit{Chandra} count rates are used only to estimate the unabsorbed luminosity, assuming a power-law spectral shape with photon index of 1.7 as was found from the low quality \textit{XMM-Newton} data (see below). 

We carried out a spectral and temporal analysis on all the available \textit{XMM-Newton} observations (Table~\ref{tab1}), which were taken in \textit{Imaging} mode, of NGC 5643 ULX1. The data sets were reduced using SAS v. 13.5.0 and the most recent EPIC calibrations. After selecting good time intervals free of background flares, we extracted spectra and lightcurves from events with {\sc pattern}$\leq 4$ for EPIC-pn (which allows for single and double pixel events) and {\sc pattern}$ \leq 12$ for EPIC-MOS (which allows for single, double, triple and quadruple pixel events). We set `{\sc flag}=0' in order to exclude bad pixels and events coming from the CCD edges. EPIC-MOS and pn source and background spectra were extracted from 20'' radius circular regions (the latter on the same CCD where the source is located and far from the read-out node). These were chosen to be smaller than is typically used for \textit{XMM-Newton} data in order to reduce contamination by the nucleus of NGC 5643, which is located $\sim50$$''$ from the ULX. 
Finally, we note that, in observation 2, NGC 5643 ULX1 was on a CCD gap of the EPIC-pn instrument, as such we only analysed the EPIC-MOS data.

All of the extracted spectra were rebinned such that they had at least 25 counts per bin so that they could be analysed using the $\chi^2$ statistic. Spectral fitting was performed using XSPEC v.12.8.2 \citep{arnaud96}. For each observation, we fitted the EPIC-pn (except for observation 2) and both EPIC-MOS spectra simultaneously in the 0.3--10 keV energy range.
In all fits, a multiplicative constant was used to account for residual calibration differences between the three instruments. The constant was fixed to 1 for EPIC-pn and allowed to vary for the EPIC-MOS spectra. In the case of observation 2, we fixed the constant to 1 of the EPIC-MOS1 spectrum.
In general, the differences among the three instruments were less than 5$\%$.

\section{Analysis of the ULX}
\label{spectral_fits}

\subsection{High quality data}

\subsubsection{Timing analysis}
We start by reporting results from observation 3, as it has the best quality data. This observation caught NGC 5643 ULX1 in a constant flux state, with an average count rate in the EPIC-pn instrument of $\sim0.3$ cts s$^{-1}$ (Figure~\ref{lc}-\textit{bottom}).

\begin{figure}
\center
\includegraphics[width=9.3cm,angle=360]{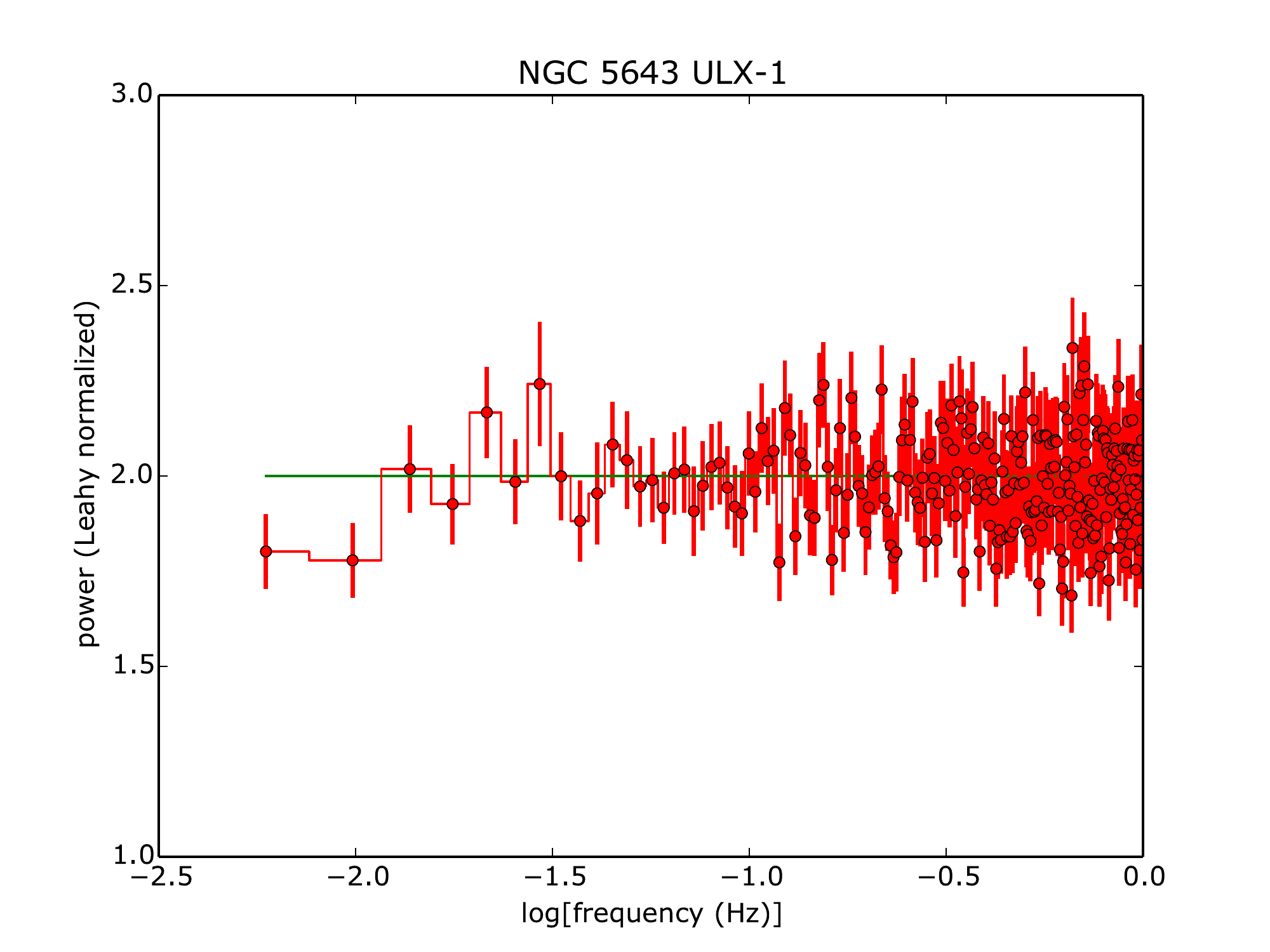}
\caption{PSD of observation 3 averaged over 10 intervals and rebinned by a factor of 32, evaluated in the frequency range $6\times10^{-3}-1$ Hz. Poisson noise (Leahy normalisation) has a constant power of 2 (green solid line). No coherent signals or significant temporal variability is found above the Poisson noise.}
\label{PDS}
\end{figure}

To check for short-term variability, we calculated the EPIC-pn power spectral density (PSD) on the timescale of the longest continuous segment in this observation (77 ks). There were no significant features in the PSD, indicating that the variability was consistent with white noise. The upper limit to root mean square (rms) variability of periodic signals in excess of the white noise, in the frequency range $1.3\times10^{-5}-5.0$ Hz, was 4.3$\%$ at the 90 per cent confidence level (c.l.). The confidence levels refers to the probability (expressed in terms of the Gaussian probability distribution) that the variability of a signal is lower than the reported threshold (see \citealt{vanderklis94}). We also calculated PSDs in the $6\times10^{-3}$--1 Hz frequency range, which were averaged over 10 intervals and linearly rebinned in frequency to 32 bins (Figure~\ref{PDS}). Again, no significant variability was detected above the Poisson noise. We calculated the upper limit to the rms of the signal in the frequency range $1.3\times10^{-4}-1$ Hz to be 10.4$\%$ at the 90.0 \% c.l.. 

Finally, because no short-term variability was detected above the Poisson noise, we also estimated the 3$\sigma$ upper limit on the rms fractional variability \citep[e.g.][]{vaughan03} to be $<$9\%, in the timescale 100s--77ks. {For a direct comparison with observation 1 (see section~\ref{low}), we also estimated the rms fractional variability on timescales 100s-8000s, finding a 3$\sigma$ upper limit of $\sim3\%$.}

\begin{table*}
\footnotesize
\begin{center}
   \caption{Best fitting spectral parameters for several different models fitted to the {\it XMM-Newton} spectra of NGC 5643 ULX1. The errors are at the 90$\%$ c.l. for each parameter.}
\scalebox{0.78}{\begin{minipage}{24cm}
   \label{table}
   \begin{tabular}{l c c c c c c c c c c }
\hline
Model &  $N_{\text{H}}$$^a$ & $kT_{\text{in}}$$^{b,c}$ & $p^d$ & $Norm_{\text{disc}}$$^e$& $\Gamma^f$ & $kT_{\text{e}}$$^{g,h}$ & $\tau$$^i$ &  $Norm.^j$ &$L_{X}$ [0.3-10 keV] $^k$ & $\chi^2/dof$$^l$ \\ 
& ($10^{21}$ cm$^2$) & (keV) & &   &  &(keV)  &  & &($10^{40}$ erg s$^{-1}$) &\\
\hline
{\bf Observation 1} &&&&&&&\\
\hline
{\sc powerlaw}& $1.1_{-0.4}^{+0.4}$	& - &- &-& $1.71_{-0.09}^{+0.1}$   &-& - & 1.8$_{-0.2}^{+0.2}\times10^{-4}$ & $4.3_{-0.3}^{+0.2}$& 102.46/93\\
{\sc diskbb}&  $<8.2$	& $1.5_{-0.1}^{+0.1}$ & - & $8_{-1}^{+2}\times10^{-3}$ & -& - &  - &-&$3.2_{-0.2}^{+0.2}$& [137.84/93]\\
{\sc diskpbb}&  $0.8_{-0.5}^{+0.5}$	& $3.6_{-1.2}^{+3.6}$ & 0.56$_{-0.03}^{+0.04}$ & $1_{-1}^{+6}\times10^{-4}$ &-&  - &  - &-&$4.0_{-0.5}^{+0.4}$& 100.86/92\\
\hline
{\bf Observation 2} &&&&&&&\\
\hline
{\sc powerlaw}& $<0.8$	& - &- &-& $1.8_{-0.1}^{+0.1}$   &-& - & 0.80$_{-0.09}^{+0.09}\times10^{-4}$ & $1.7_{-0.1}^{+0.2}$& 21.65/22\\
{\sc diskbb}&  $<8.2$	& $1.1_{-0.2}^{+0.2}$ & - & $1.1_{-0.4}^{+0.7}\times10^{-2}$ & -& - &  - &-&$1.2_{-0.2}^{+0.1}$& [50.99/22]\\
{\sc diskpbb}&  $<0.7$	& $9.8_{-7.7}^{+8.0}$ & 0.53$_{-0.03}^{+0.02}$ & $7_{-3}^{+34}\times10^{-7}$ &-&  - &  - &-&$1.6_{-0.2}^{+0.3}$& 21.18/21\\
\hline
{\bf Observation 3} &&&&&&&\\
\hline
{\sc powerlaw}& $1.14_{-0.08}^{+0.08}$	& - &- &-& $1.70_{-0.02}^{+0.02}$   &-& - & 1.92$_{-0.05}^{+0.05}\times10^{-4}$ & $4.3_{-0.1}^{+0.1}$& 1001.86/936\\
{\sc diskbb}&  $<4e^{-3}$	& $1.60_{-0.03}^{+0.03}$ & - & $7.71_{-0.07}^{+0.07}\times10^{-3}$ & -& - &  - &-&$3.28_{-0.04}^{+0.04}$& [1220.74/936]\\
{\sc diskbb+powerlaw}& $2.49_{-0.06}^{+0.06}$ & $0.130_{-0.007}^{+0.009}$ &- &$230_{-140}^{+279}$ & $1.79_{-0.04}^{+0.04}$   & -& - & 2.23$_{-0.01}^{+0.01}\times10^{-4}$ & $6.2_{-0.5}^{+0.4}$& 981.16/934\\
{\sc powerlaw+diskbb}& $1.1_{-0.4}^{+0.5}$ & $2.1_{-0.2}^{+0.1}$ &- &$1.6_{-0.3}^{+0.4} \times 10^{-3}$ & $2.2_{-0.3}^{+0.5}$   & -& - & 1.2$_{-0.1}^{+0.2}\times10^{-4}$ & $4.4_{-0.2}^{+0.6}$& 884.41/934\\
{\sc cut-off powerlaw}& $0.4_{-0.1}^{+0.1}$	& - &- &-& $1.1_{-0.1}^{+0.1}$   &5.4$_{-0.8}^{+1.0}$& - & 1.81$_{-0.05}^{+0.05}\times10^{-4}$ & $3.9_{-0.1}^{+0.1}$& 892.41/935\\
{\sc diskpbb}&  $0.6_{-0.1}^{+0.1}$	& $2.6_{-0.2}^{+0.2}$ & 0.60$_{-0.01}^{+0.01}$ & $6_{-2}^{+2}\times10^{-4}$ &-&  - &  - &-&$3.8_{-0.1}^{+0.1}$& 886.13/935\\
{\sc comptt}& $0.8_{-0.2}^{+0.1}$& $0.10_{-0.03}^{+0.03}$ &- & - &  -&  $1.73_{-0.02}^{+0.02}$ &$9.7_{-0.5}^{+0.6}$ &3.8$_{-0.5}^{+1}\times10^{-4}$ & $3.8_{-0.1}^{+0.2}$& 877.57/934\\
{\sc diskbb+comptt}& $<0.3$& $0.26_{-0.04}^{+0.08}$ &- & 1.0$_{-0.2}^{+1.0}$  &-&  $1.61_{-0.08}^{+0.08}$ &  $10.9_{-0.5}^{+0.9}$ &2.1$_{-0.3}^{+0.3}\times10^{-4}$ & $3.44_{-0.06}^{+0.12}$& 875.41/933\\
\hline
   \end{tabular} 
\end{minipage}}
 \end{center}
\begin{flushleft} $^a$ Intrinsic column density in excess to the Galactic one; $^b$ inner disc temperature; $^c$ seed photons temperature for Comptonisation; $^d$ exponent of the radial dependence of the disc temperature; $^e$ normalisation of the disc component; $^f$ photon index of the {\sc powerlaw} component; $^g$ electron temperature of the corona; $^h$ e-folding energy of the {\sc cutoffpl} model; $^i$ optical depth of the corona; $^j$ normalisation of the {\sc comptt} or {\sc powerlaw} component; $^k$ unabsorbed total X-ray luminosity in the 0.3--10 keV energy band assuming a distance of 16.9 Mpc; $^l$ reduced $\chi^2$; square brackets indicate that the fit is not statistically acceptable.\\
\end{flushleft}
\end{table*}

\begin{figure*}
\center
\subfigure{\includegraphics[width=6.5cm,height=8.5cm,angle=-90]{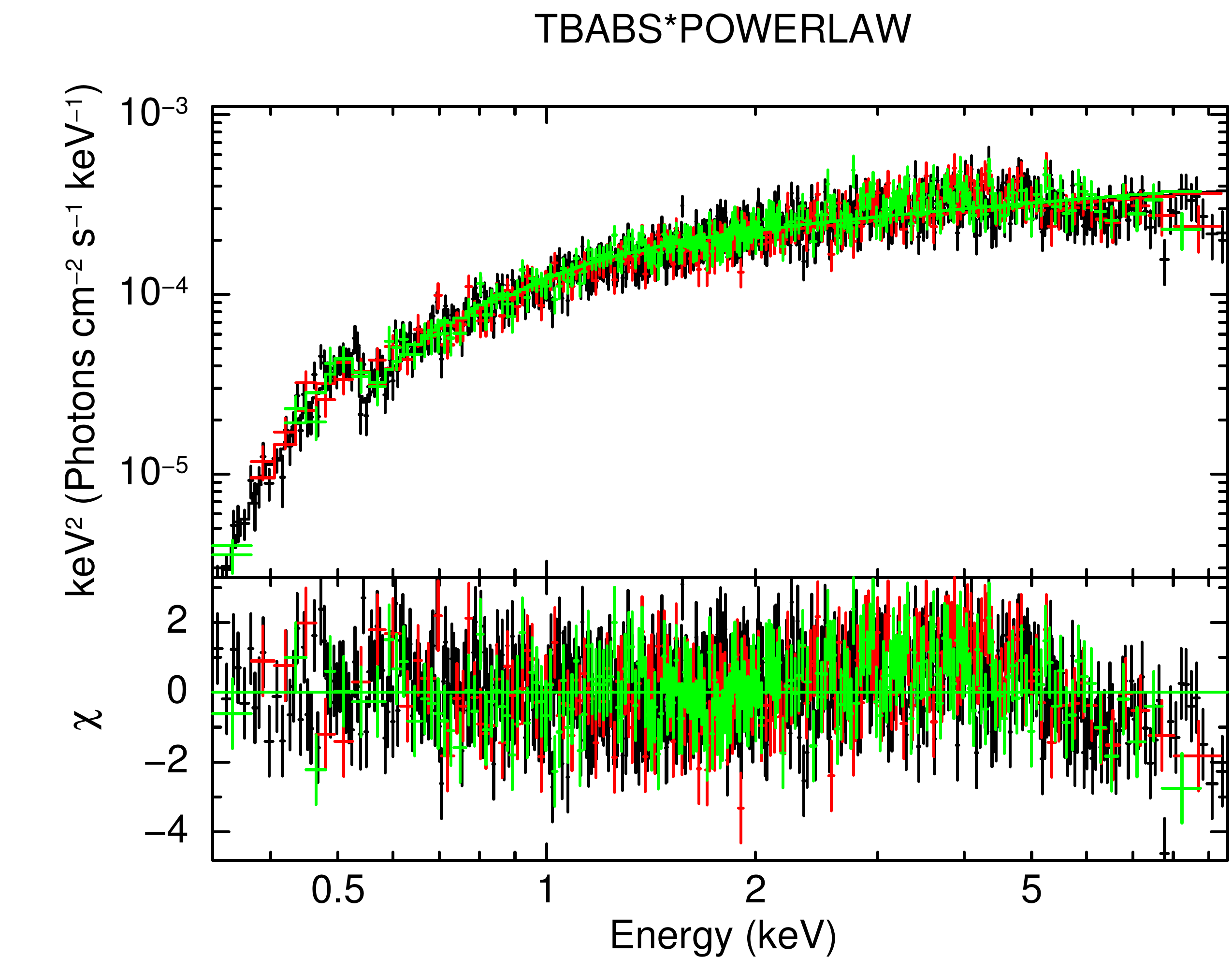}}
\hspace{3mm} \subfigure{\includegraphics[width=6.5cm,height=8.5cm,angle=-90]{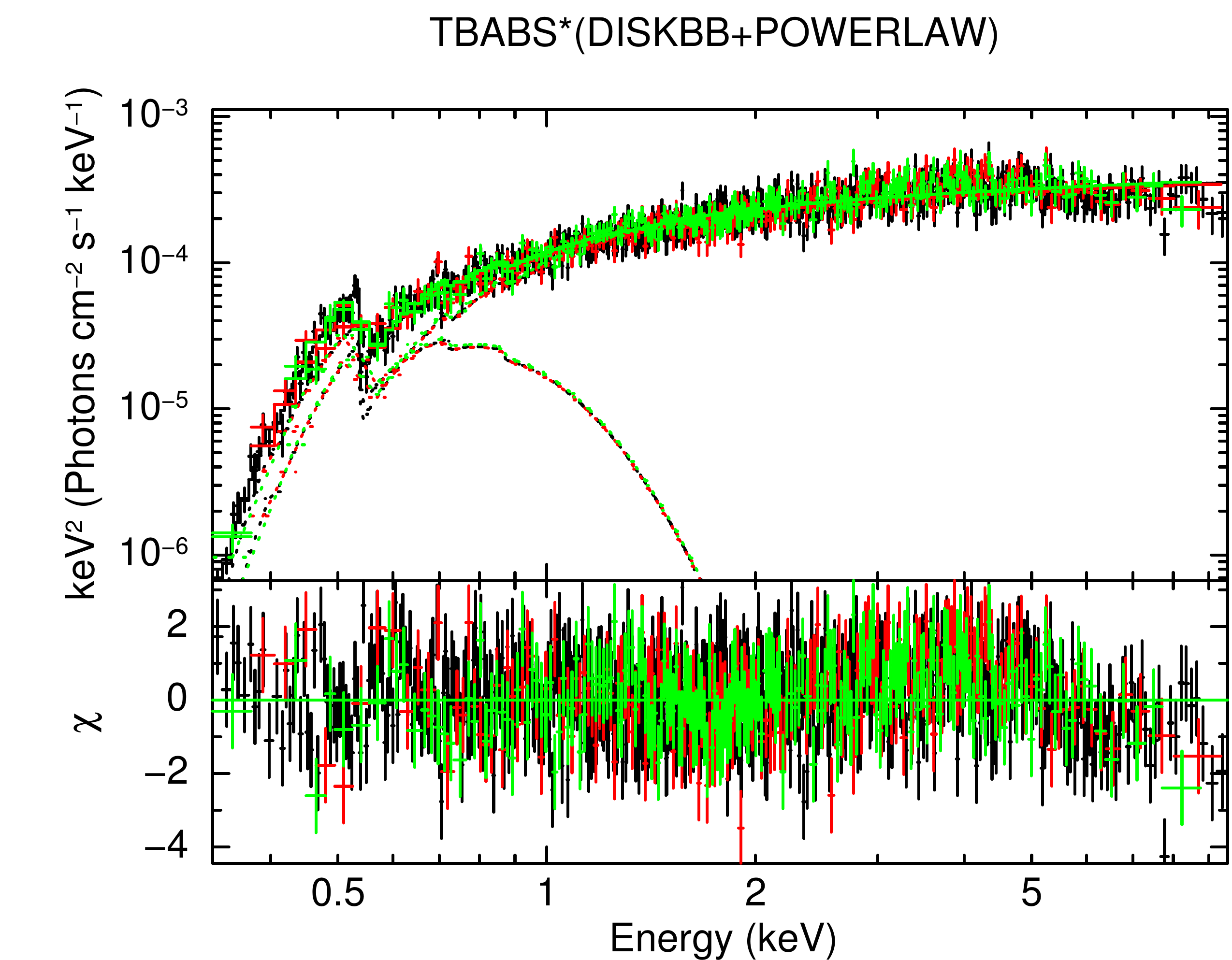}}
\subfigure{\includegraphics[width=6.5cm,height=8.5cm,angle=-90]{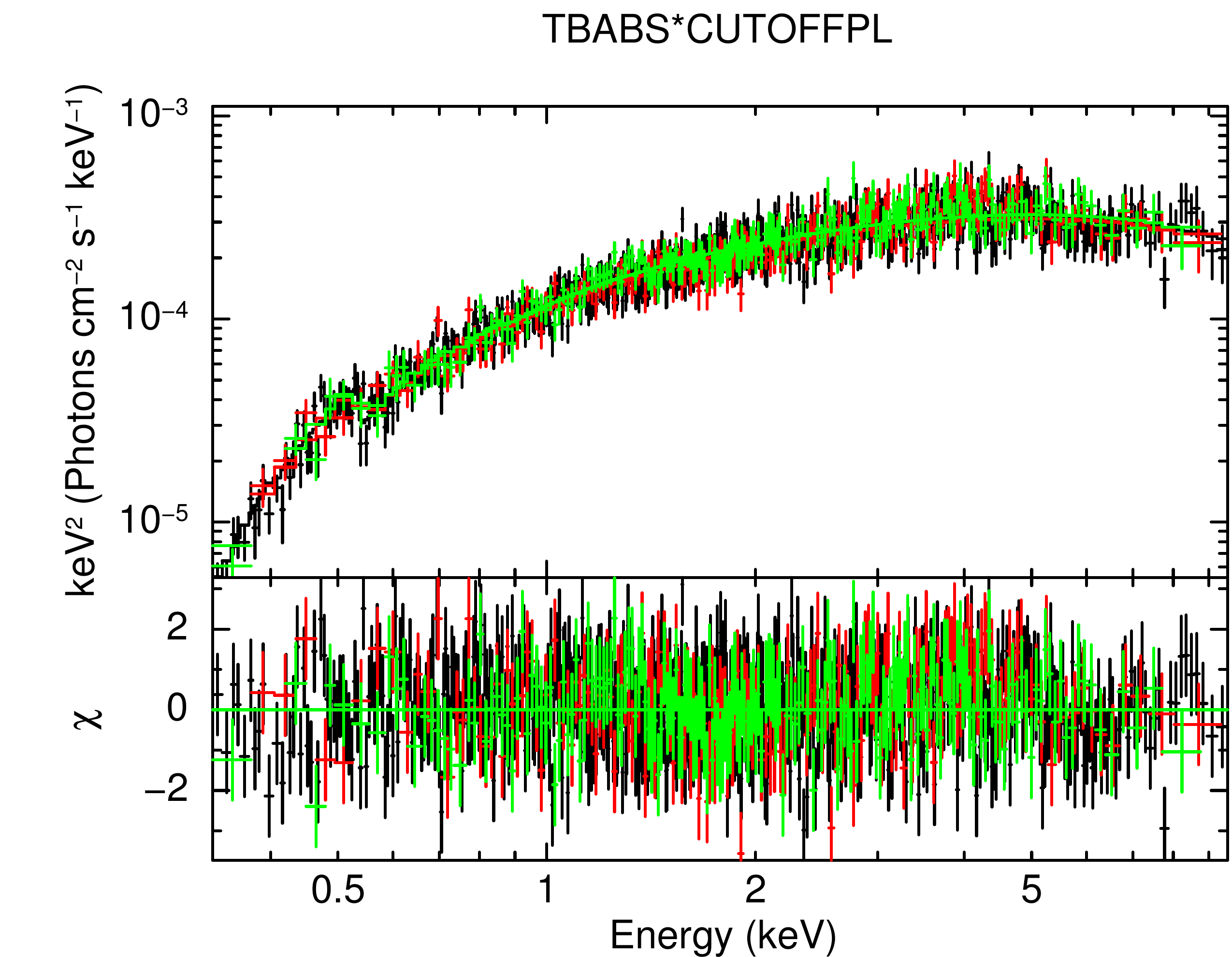}}
\hspace{3mm}\subfigure{\includegraphics[width=6.5cm,height=8.5cm,angle=-90]{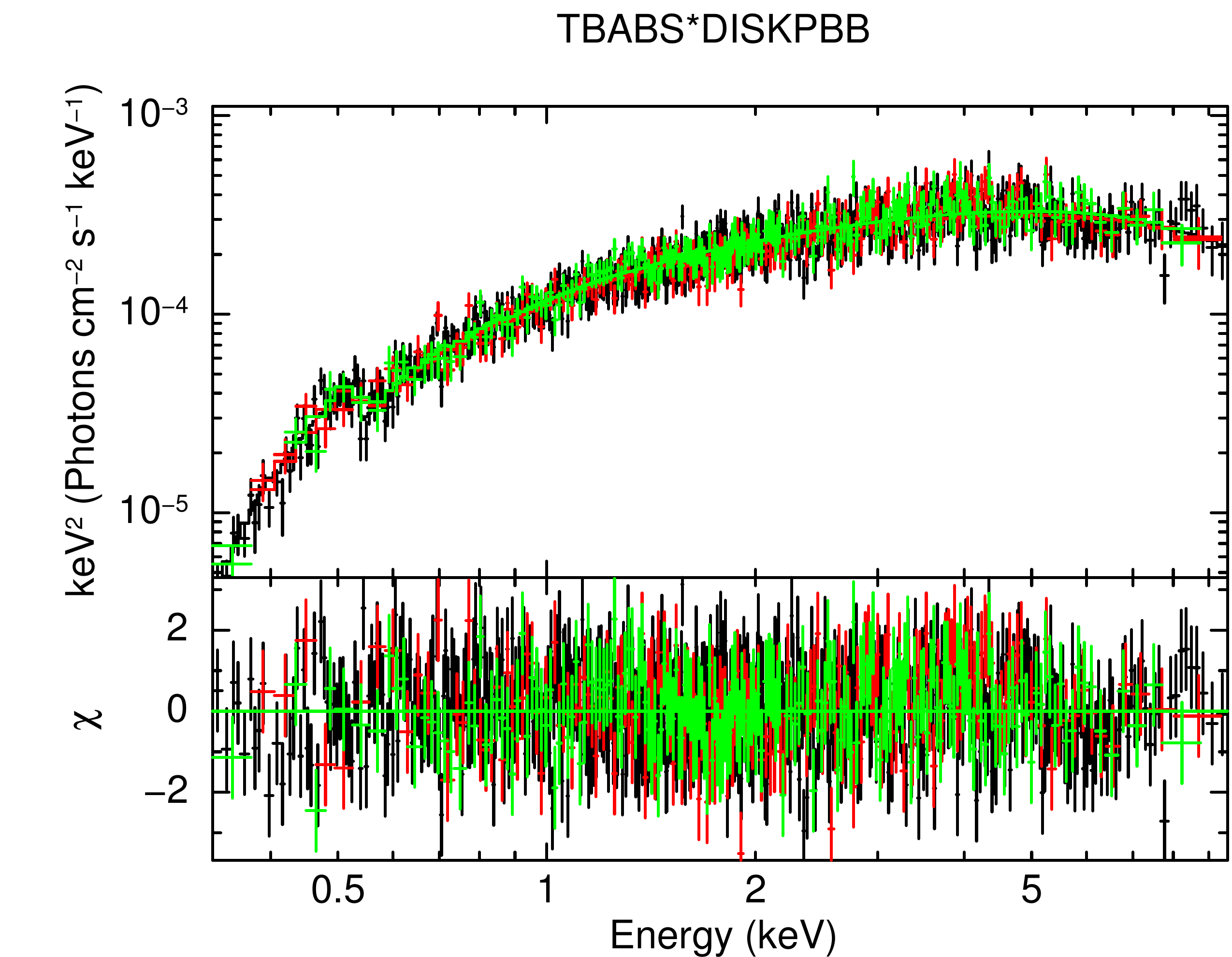}}
\subfigure{\includegraphics[width=6.5cm,height=8.5cm,angle=-90]{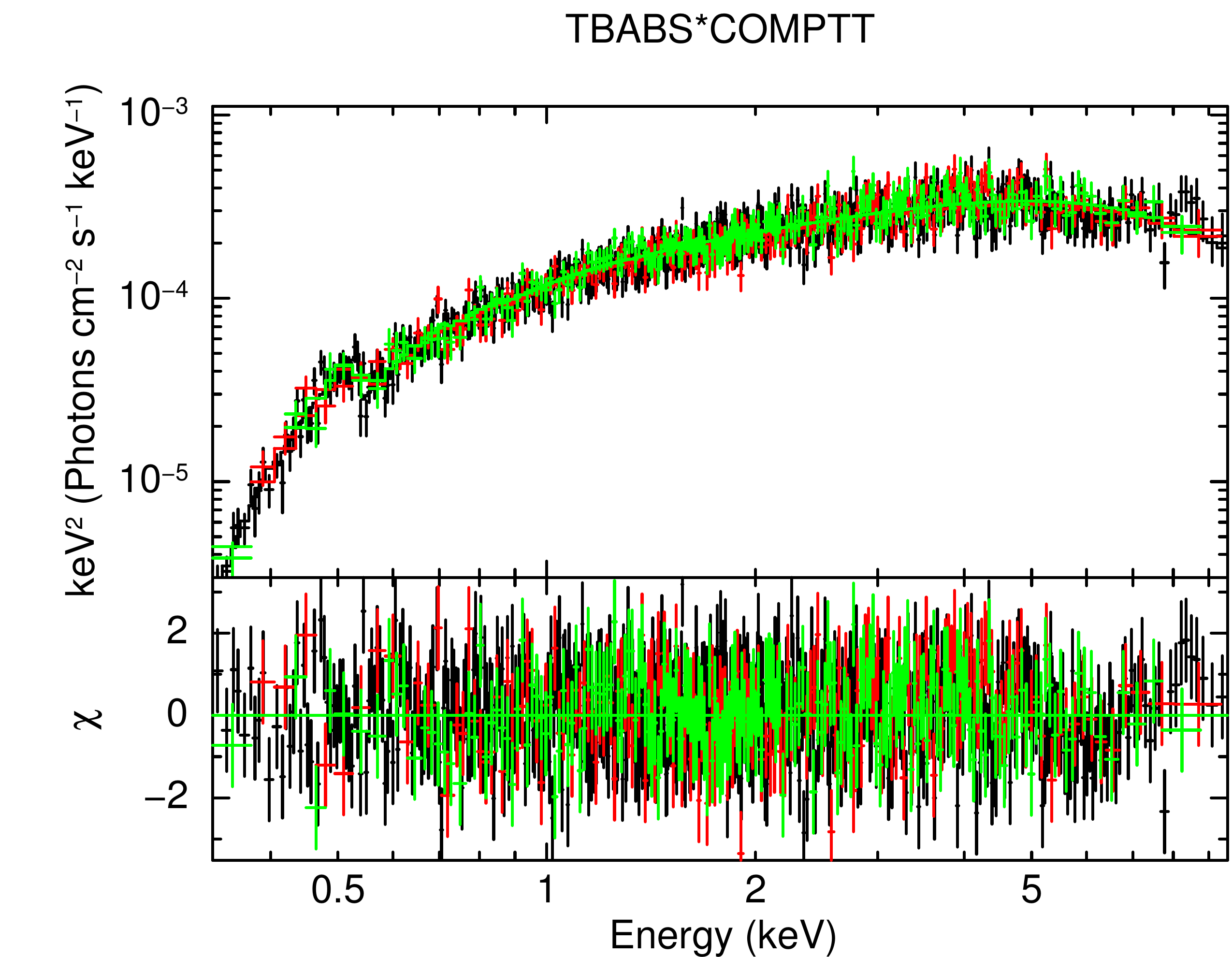}}
\hspace{3mm}\subfigure{\includegraphics[width=6.5cm,height=8.5cm,angle=-90]{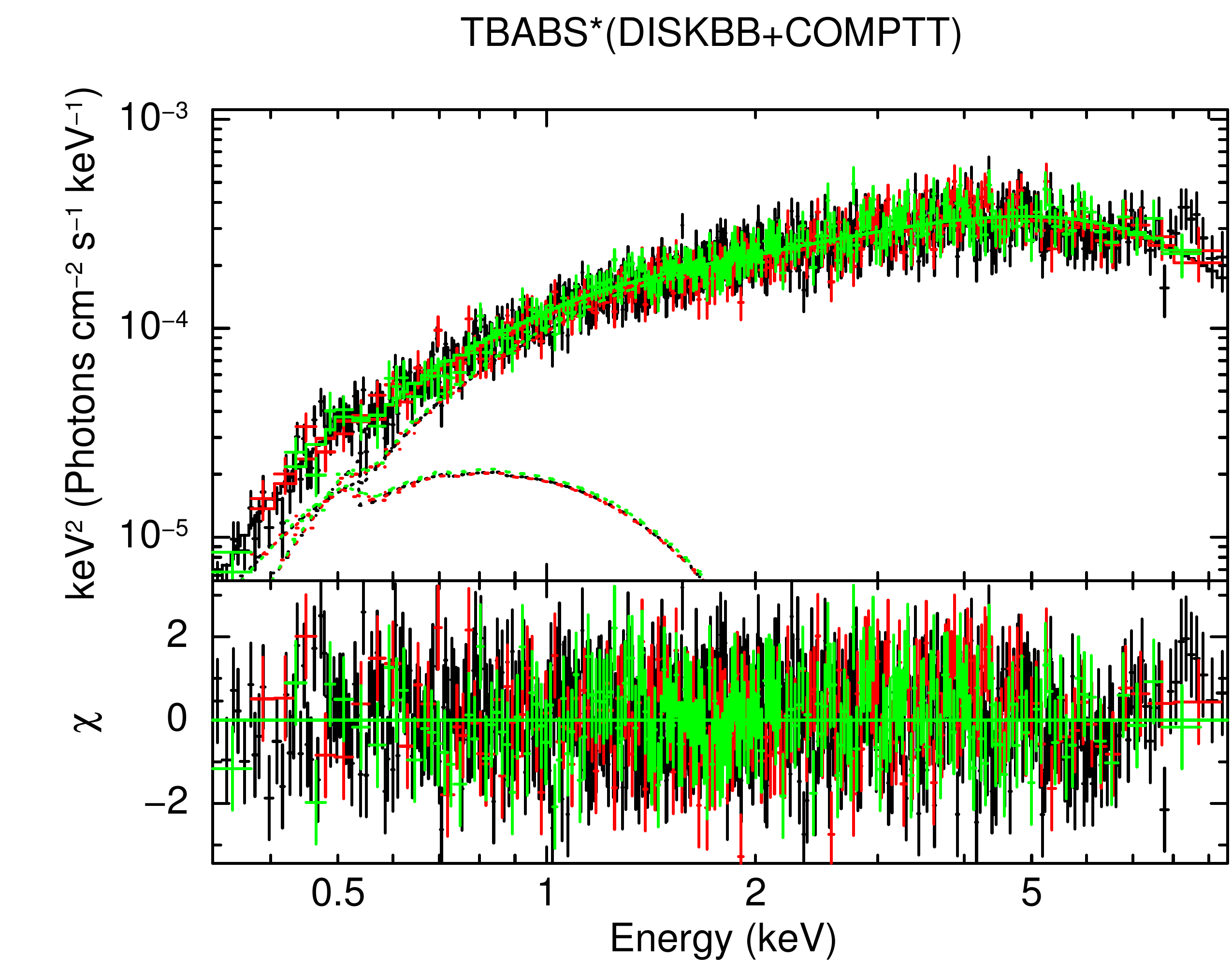}}
\caption{EPIC-pn (black) and EPIC-MOS (red and green) spectra of observation 3 unfolded from the detector response, with the particular models indicated by the title of each plot. The power-law  model clearly resulted in large residuals above 5 keV, but the spectra could be well modelled by either an advection dominated disc ({\sc diskpbb}) or a single Comptonisation model ({\sc comptt}).}
\label{models}
\end{figure*}

\subsubsection{Spectral analysis} 
The first stage of our spectral analysis was to look for short-term spectral variability during the long observation. We extracted light curves with 500s binning in the 0.3--1.5 and 1.5--10 keV energy ranges, which were selected such that there were similar count rates in each band. The ratios of the corresponding light curves are consistent with being constant ($\sim1$) at 3$\sigma$ significance. As such, we analyse the time-averaged X-ray spectrum from the entire observation.

We initially fitted the EPIC-pn and EPIC-MOS spectra with a power-law absorbed by neutral material. We considered two neutral absorbers (modeled with {\sc tbabs} in {\sc XSPEC}), one fixed to the Galactic value in the direction of the ULX \citep[$8\times10^{20}$ cm$^{-2}$; ][]{kalberla05}, and the other free to vary to model absorption close to the ULX. Although this resulted in a statistically acceptable fit ($\chi^2$ of 1002 for 936 degrees of freedom, with parameters given in Table~\ref{table}), the power-law model resulted in  hard residuals. Using the {\sc cflux} convolution pseudo-model we estimated the unabsorbed X-ray luminosity to be $\sim4\times10^{40}$ erg s$^{-1}$ (for a distance of 16.9 Mpc). 
The large fit statistic is mainly due to a clear roll-over at energies higher than 5 keV that cannot be reproduced by a power-law (see the residuals in Figure~\ref{models}-\textit{top-left}). 
To test the statistical significance of the high energy roll-over we followed the method of \citet{stobbart06}, and compared a power-law and a broken power-law ({\sc bknpower} in {\sc xspec}) model fitted to the $> 2$ keV spectra. We found that a {\sc bknpower} fit with the two photon indexes tied together is statistically acceptable ($\chi^2$/dof = 506.31/423) but a significant improvement in the fit ($\Delta\chi^2$ = 103.54 for 2 additional dof) is found by allowing them to vary. The broken power-law fit provided a clearly superior description, with a F-test probability of $< 3\times10^{-18}$ that the improvement was due to a random fluctuations in an intrinsic power-law spectrum.

In the sub-Eddington {\it hard} state XRBs exhibit both high levels of short-term variability (larger than 10$\%$) and a hard power-law spectrum \citep[e.g.][]{mcclintock06}. 
As such, the detection of a high energy break and the absence of short term variability allow us to robustly conclude that NGC 5643 ULX1 was not in the sub-Eddington \textit{hard} state during observation 3. 
To test whether it may have been in the sub-Eddington {\it soft} state we also fitted the spectra with an absorbed multicolour disc ({\sc diskbb}; \citealt{mitsuda84}), finding that this resulted in a statistically poorer fit than the power-law ($\chi^2/dof=1221/936$). This excludes the possibility that ULX1 was in the sub-Eddington {\it soft} state.

Next, we considered a multicolour disc added to the power-law model. Adopting this two-component model, we found a good fit with a hot disc ($\sim2$ keV) with the power-law dominating at low energies ($\Delta\chi^2=117.45$ for 2 additional dof with respect to a single power-law; see Table ~\ref{table}). Additionally, a cool disc ($0.13$ keV) and a {powerlaw} results in an improved fit over a power-law alone, although this is at lower significance than the hot disc ($\Delta\chi^2=20.7$ for 2 additional dof, $\chi^2$/dof$=981/934$). 
Although this is statistically acceptable, residuals can be clearly seen at energies $> 5$ keV (Figure~\ref{models}-\textit{top-right}). Furthermore, when combined with the cool disc the power-law has a photon index of $\Gamma\sim1.8$, which is inconsistent with the {\it steep powerlaw} state seen in Galactic BHs ($\Gamma>2.2$). 

Based on the spectral analysis of observation 3, we are able to conclude that NGC 5643 ULX1 was
not in any of the canonical sub-Eddington spectral states seen in Galactic BHs. 
Hence, based on the assumption that sub-Eddington IMBHs would be in the same accretion states as Galactic XRBs (albeit with some rescaling to account for the BH mass), we may strongly rule-out the possibility that NGC 5643 ULX1 contains an IMBH. 
Furthermore, the degeneracy between hot and cool accretion discs when fitting the spectra with the two-component multicolour disc plus power-law model is not unusual in ULXs and it has been investigated in the past \citep[e.g.][]{stobbart06}. Therefore, the spectral properties of NGC 5643 ULX1 appear to be consistent with the standard ULX population. 
As such, we attempted to classify NGC 5643 ULX1 using the multicolour disc plus power-law model, according to the scheme proposed by \citet{sutton13}.
Since we found two statistical minima for this model (i.e. a hot or a cold disc model), we may classify the source as being in either the {\it broadened-disc} state or the {\it hard-ultraluminous} state. However, the minimum with a hot disc was statistically preferable, suggesting that NGC 5643 ULX1 is best classified as a {\it broadened-disc} source (although see Section~\ref{discussion} for caveats with this interpretation).

To further investigate the spectral properties of the source, particularly the high energy cut-off, we adopted several more spectral models.
The first of these was a single power-law with an exponential cut-off ({\sc cutoffpl} in {\sc xspec}). This model strongly improved the fit statistic compared to the standard power-law ($\Delta\chi^2$$=109.45$ for 1 additional dof, Figure~\ref{models}-\textit{centre-left}) and strongly reduced the high energy residuals. The best fitting model parameters were an intrinsic neutral hydrogen column density of 4$\times10^{20}$ cm$^{-2}$ and a power-law photon index of 1.1, which broke at $\sim5.5$ keV. Similar fits are not unusual for ULXs \citep[e.g.][]{gladstone09}. However, this model is phenomenological. In terms of more physical models, the spectrum of NGC 5643 ULX1 can be well described by both an advection-dominated slim disc ({\sc diskpbb}\footnote{This model describes a modified accretion disc with  $T(r) \propto r^{-p}$ ($r$ is the disc radius), with $p=0.75$ corresponding to a standard \citep{shakura73} disc, and $p=0.5$ to an advection dominated slim disc \citep[e.g.][]{kawaguchi03}.}; e.g. \citealt{mineshige94}) and a Comptonisation model ({\sc comptt}; \citealt{titarchuk94}). Both models are statistically acceptable (see Table~\ref{table}) and show that a broad, thermal-like component was a good description of the entire 0.3--10 keV spectrum. 

\begin{figure*}
\center
\subfigure{\includegraphics[width=6.5cm,height=8.5cm,angle=-90]{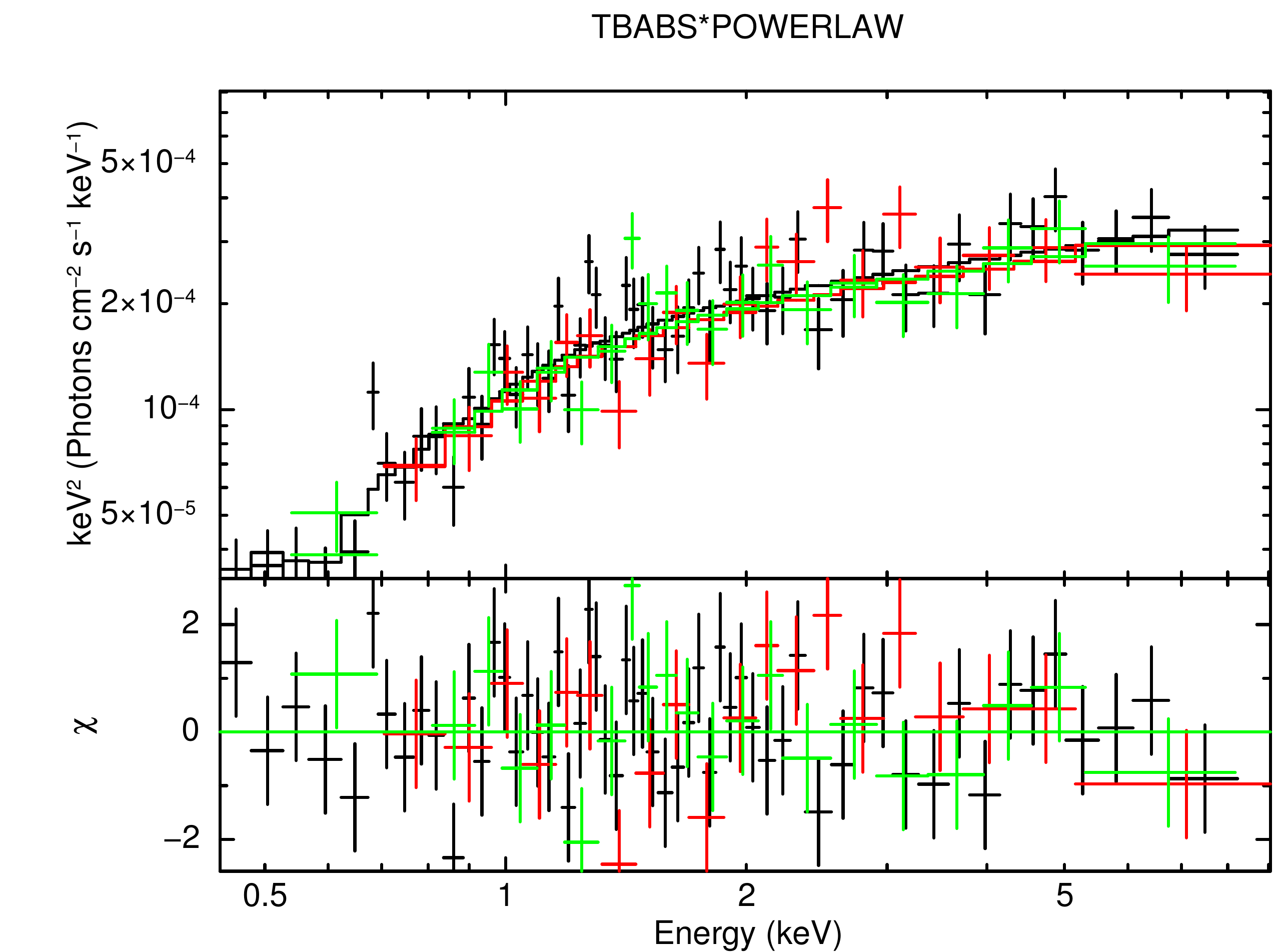}}
\hspace{3mm} \subfigure{\includegraphics[width=6.5cm,height=8.5cm,angle=-90]{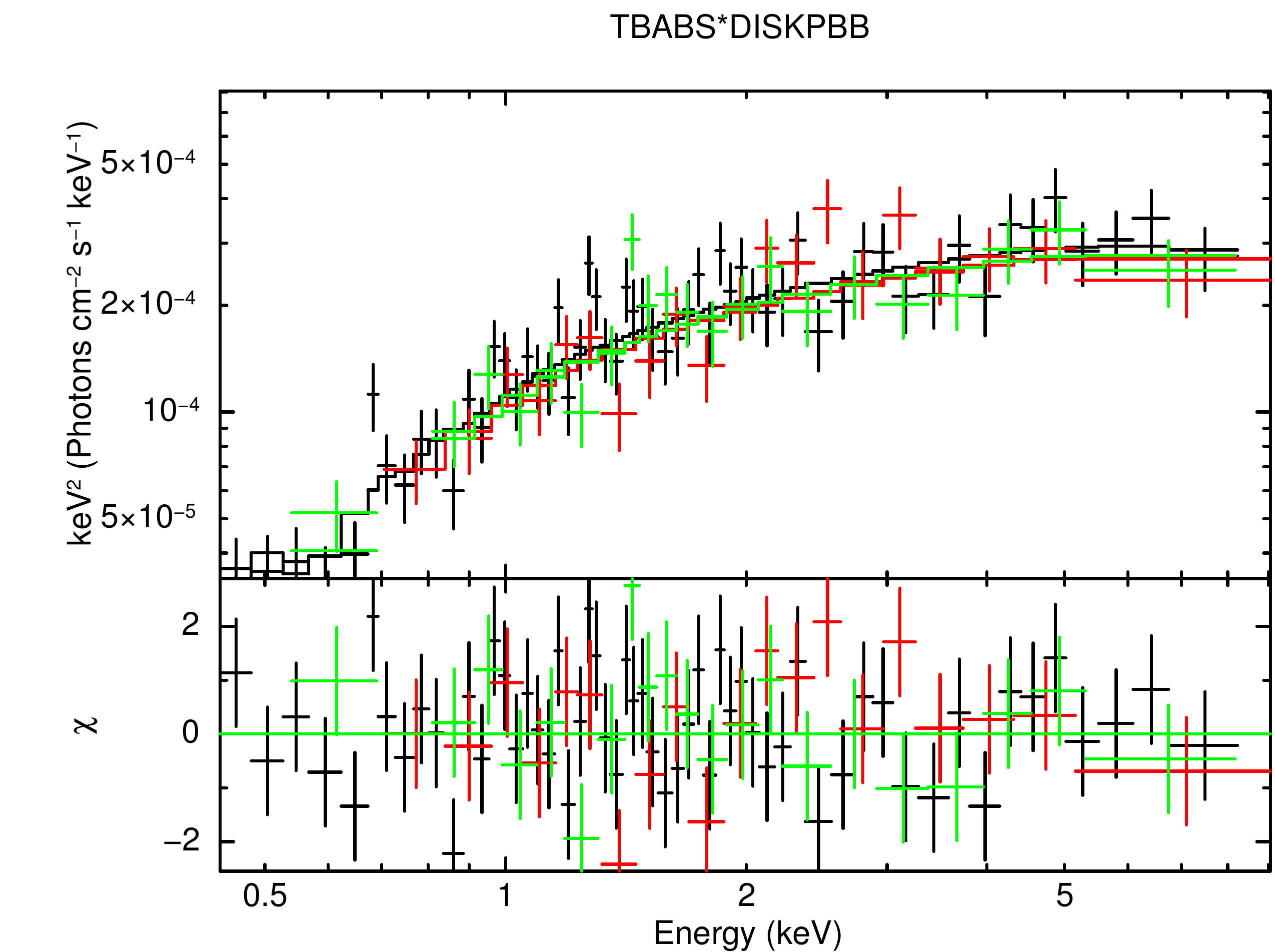}}
\subfigure{\includegraphics[width=6.5cm,height=8.5cm,angle=-90]{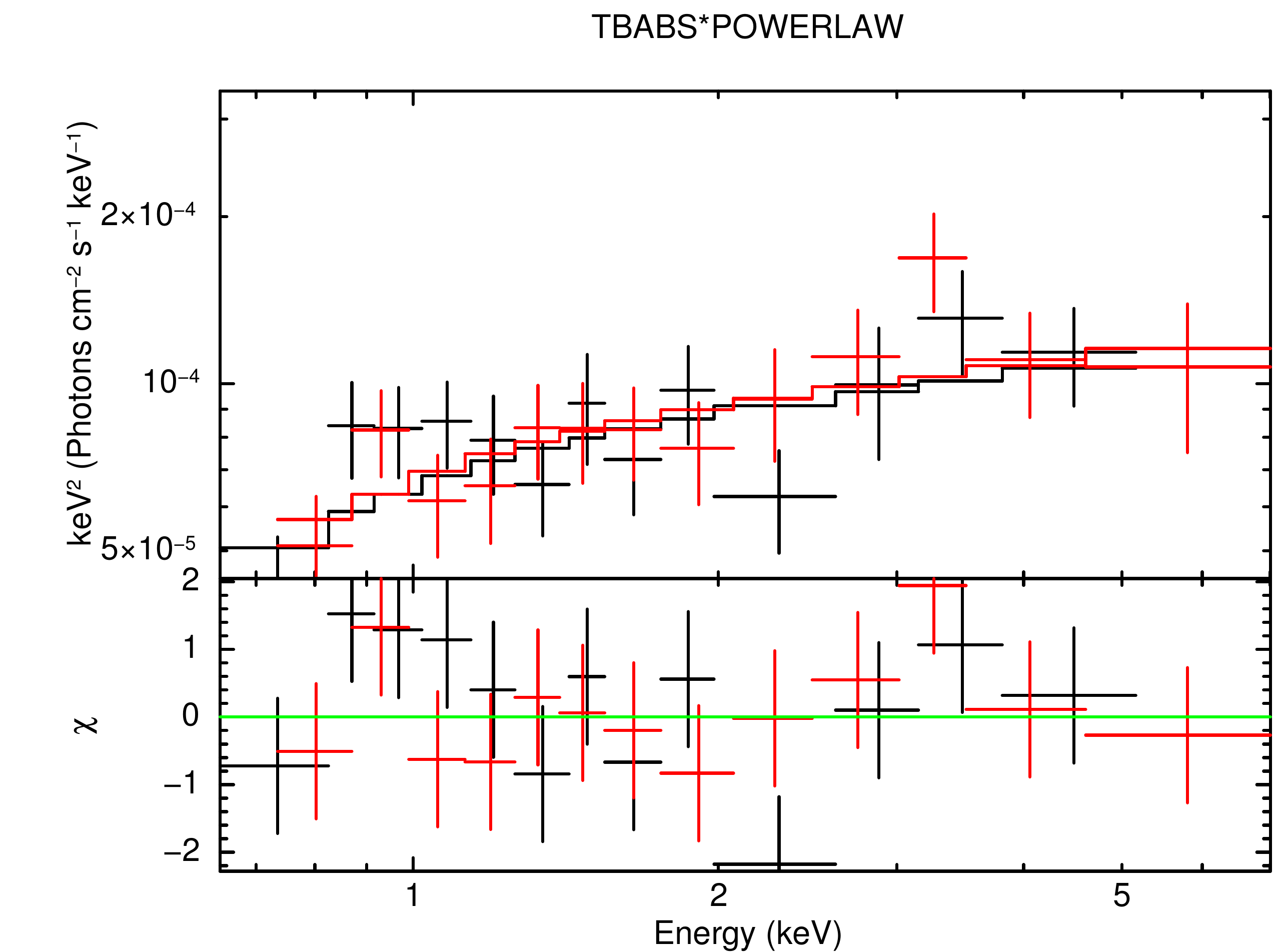}}
\hspace{3mm} \subfigure{\includegraphics[width=6.5cm,height=8.5cm,angle=-90]{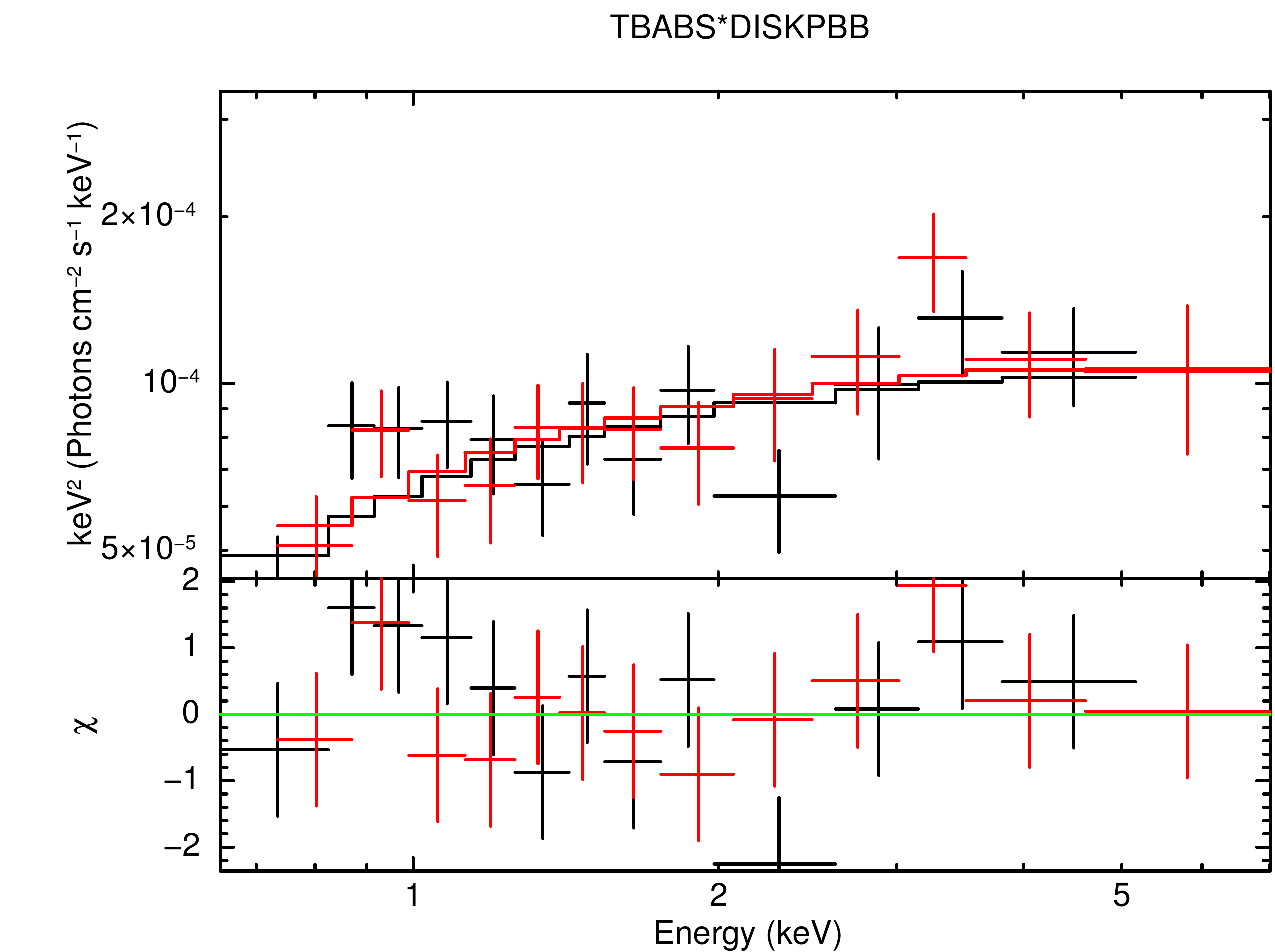}}
\caption{EPIC-pn (black) and EPIC-MOS (red and green) spectra extracted from observation 1 (\textit{top}) and 2 (\textit{bottom}) unfolded from the detector response, with the particular models indicated by the title of each plot.}
\label{models_12}
\end{figure*}

The best fitting parameters for the {\sc diskpbb} model gave an intrinsic neutral column density of $6\times10^{20}$ cm$^{-2}$ and a disc temperature of 2.6 keV. However, this temperature is clearly high and unusual with respect to those observed in Galactic BHs \citep{mcclintock06}. Additionally, the value of the \textit{p} parameter suggests that the disc has to be significantly advection dominated (p=$0.60\pm0.01$) and hence that the accretion rate in the disc is likely super-Eddington. From this model we infer a 0.3--10 keV unabsorbed luminosity of $3.8\times10^{40}$ erg s$^{-1}$ (flux of $1.2\times10^{-12}$ erg s$^{-1}$ cm$^{-2}$).
Alternatively, the Comptonisation model resulted in a column density of $8\times10^{20}$ cm$^{-2}$ (consistent with the previous model within the uncertainties) and corresponded to a cold, optically thick corona ($\tau\sim10$ and kT$_{e}\sim2$ keV), fed by seed photons with a temperature of 0.1 keV. The estimated unabsorbed luminosity is consistent with that found with the {\sc diskpbb} model. We note that a few 0.3--10 keV spectra of high luminosity ULXs ($> 10^{40}$ erg s$^{-1}$) can be also described by only a single component model, such as Comptoniation or an advection dominated disc (e.g. IC 342 X-1, \citealt{gladstone09}, although see \citealt{sutton13} for caveats about this interpretation).

It has been suggested that the hard spectral component observed in ULXs may originate from an optically thick corona above the accretion disc, or from the inner disc itself \citep[e.g.][]{stobbart06,gladstone09}. High quality X-ray spectra of luminous ULXs can often be well-fitted by a two-component model consisting of Comptonisation and another soft component. Here, the soft component is associated with an optically thick wind ejected by the disc when the accretion rate locally exceeds the Eddington limit \citep[e.g.][]{poutanen07,ohsuga09,takeuchi14}. The Comptonisation component originates from the hot, inner accretion disc \citep[e.g.][]{middleton11,kajava12,middleton15}. However, if both components are present in NGC 5643 ULX1 and the inclination angle is suitable (see e.g. \citealt{middleton15}), then we would expect to observe both of these features in its spectrum. Hence, although it is not required by the data, we tentatively added a multicolour disc component to the Comptonisation model. This does not result in a statistically significant improvement ($\Delta\chi^2\sim$ 2 for 2 additional dof), and the parameters of the corona are consistent with those derived from the  single component model. However, the temperature of the soft component ($\sim0.3$ keV) is in line with what found in many other ULXs \citep[e.g.][]{stobbart06,gladstone09,sutton13,pintore14,middleton14,walton13,bachetti14}. We note instead that the column density is poorly constrained and that the soft component is very weak and represents only a marginal fraction of the total flux ($\sim6\%$).

Finally, we highlight that irrespective of the adopted model, small structured residuals are still observed at the energies of 6--7 keV and 8 keV. The former is probably caused by a small, residual contamination from the nearby AGN (which has a strong iron emission complex; see Appendix~\ref{AGN}) in both source spectrum and background. 

\subsection{Low quality data}
\label{low}

The spectra extracted from observations 1 and 2 have only moderate numbers of counts in comparison to observation 3. They can only be used to constrain simple models, and do not provide any further significant evidence for high energy spectral curvature.

The energy spectrum extracted from observation 1 can be well described by either an absorbed power-law ($\chi^2/$dof=102.46/93) or a {\sc diskpbb} ($\chi^2/$dof=100.86/92; see Table~\ref{table} and Figure~\ref{models_12}-\textit{top}), although a multicolour disc is statistically rejected at 3$\sigma$ significance ($\chi^2/$dof=137.84/93). A Comptonisation model provides an adequate fit to the data ($\chi^2/$dof=100.76/98), but most of the spectral parameters (column density, electron temperature and seed photons temperature) are unconstrained. The single {powerlaw} results in a photon index of 1.7, and a neutral column of $1.1\times10^{21}$ cm$^{-2}$. The {\sc diskpbb} model indicates a very hot inner disc temperature ($3.6$ keV), which is consistent  with that found for observation 3. Similarly, the $p$-index ($0.56_{-0.03}^{+0.04}$) suggests an advection dominated disc, indicating that the source is accreting at above the Eddington limit. Finally, both the neutral absorption ($8\times10^{20}$ cm$^{-2}$) and unabsorbed luminosity ($\sim4\times10^{40}$ erg s$^{-1}$) are consistent with observation 3.

Observation 1 did not contain sufficient counts to calculate a meaningful PDS, although we were able to calculate the rms fractional variability. This was done in the time domain, with lightcurves binned to 100 s in order to accrue sufficient counts in each bin. We estimated the 0.3--10 keV fractional variability to be $15\pm4\%$ (on timescales of 100s--8000s, {i.e. where the duration of 8000s refers to the longest continuous segment of the observation}). This value is higher than the upper-limit from observation 3 but they are both consistent within the 3$\sigma$ uncertainties. Furthermore, despite the poor quality data of observation 1, the fluxes of observations 1 and 3 appear to be comparable (Figure~\ref{lc}-\textit{top-left}) and their spectra broadly similar.

The observation 2 spectrum (Figure~\ref{lc}-\textit{top-right}) is also well fitted by a single absorbed power-law model ($\chi^2/$dof=21.65/22; $\Gamma\sim1.7$, N$_H<8\times10^{20}$ cm$^{-2}$). This has an unabsorbed luminosity of $\sim1.7\times10^{40}$ erg s$^{-1}$, hence represents a significant drop in flux from the other observations. Again, the spectrum can also be well fitted by a {\sc diskpbb} ($\chi^2/$dof=21.18/21; see Table~\ref{table} and Figure~\ref{models_12}-\textit{bottom}), although some of the parameters are poorly constrained. As with observation 1, a multicolour disc can be strongly rejected ($\chi^2/$dof=50.99/22). {For completeness, we also investigated whether the change in luminosity could be dependent on the column density. To test this, we adopted the {\sc diskpbb} model and fixed N$_\text{H}$ to $0.6\times10^{21}$ cm$^{-2}$ (i.e. the best fit value of the {\sc diskpbb} model when fitted to the spectrum from observations 3). This test confirmed that the factor of 2-3 luminosity variability was not driven by changes in the absorption column.} 
By stacking the MOS1 and MOS2 lightcurves we were only able to place an upper limit on the 0.3--10 keV rms fractional variability (on timescales 100s--12800s) of $\sim27\%$ (estimated with the \textit{ftools} {\sc lcstats}). 

We followed two approaches to test where the observation 1 and 2 spectra were consistent with that from observation 3. 
Firstly, we simultaneously fitted the EPIC-pn (where available) and EPIC-MOS spectra from all three of the observations with either a {\sc diskpbb} or a {\sc comptt} model. The spectral parameters were fixed to the best-fitting values from observation 3, and only normalisation was free to vary between the observations. This resulted in a statistically acceptable fit ($\chi^2/$dof=623.49/637), strongly indicating that the spectral properties of the ULX remained unchanged between the observations. {In addition, we note that the normalisation of the model in observations 2 is, as previously found, more than a factor of 2 lower than those in observation 1 and 3, confirming the existence of flux variability between the datasets.}
The second approach was to create a simulated spectrum based on the parameters of the best-fitting model from observation 3. The simulated exposure time was chosen such that it was comparable to observations 1 and 2. The spectrum was then fitted with an absorbed power-law, obtaining a statistically acceptable fit, with a spectral index that was indistinguishable from observations 1 and 2 ($\Gamma\sim$1.7). 
Based on both of these tests we can conclude that the the source does not display long-term spectral variability, despite changing in its X-ray luminosity.

\section{Discussion}
\label{discussion}

\begin{figure}
\center
\includegraphics[width=8.5cm,height=6.5cm]{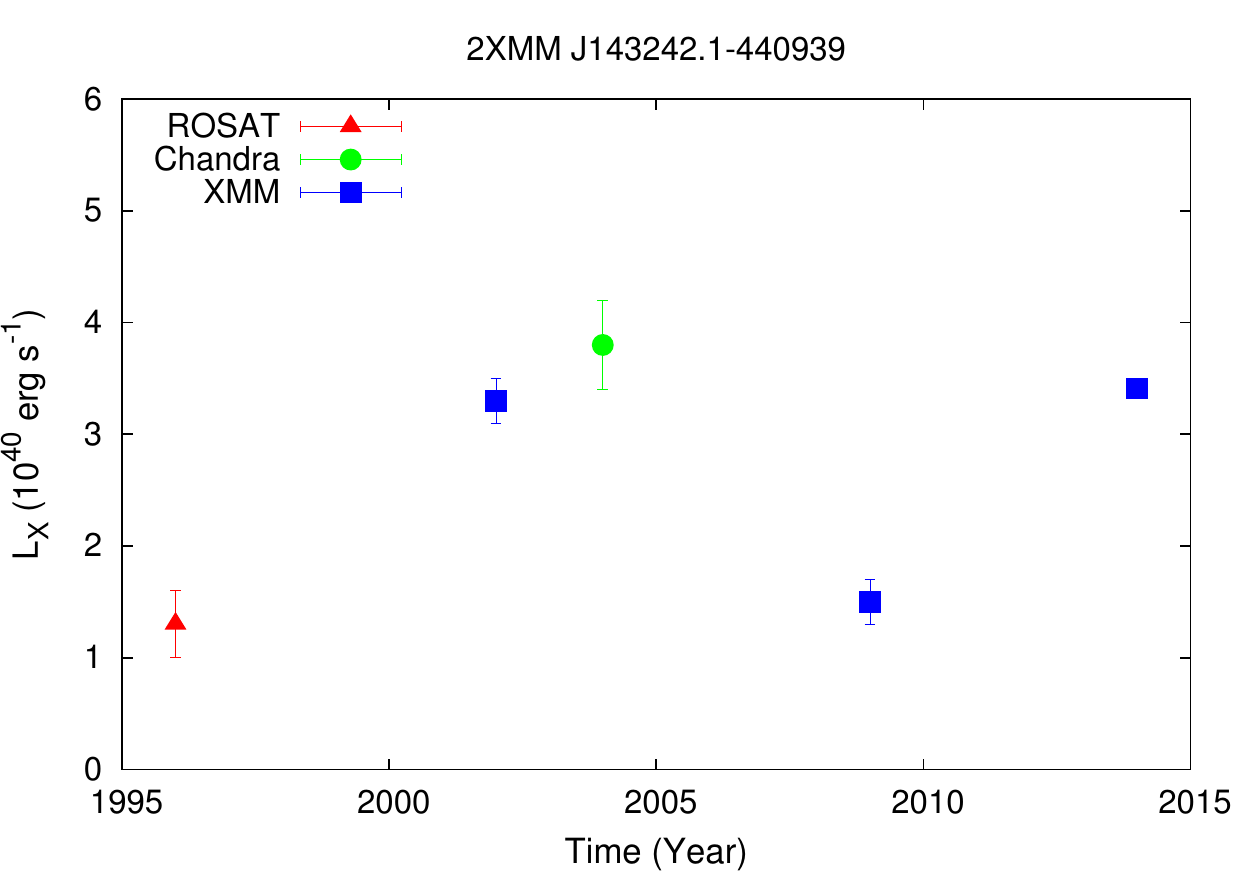}
\caption{Long-term lightcurve showing the unabsorbed luminosity of NGC 5643 ULX1 extracted from \textit{ROSAT, Chandra} and \textit{XMM-Newton} observations. Luminosities were calculated assuming a distance of 16.9 Mpc.}
\label{lc_long}
\end{figure}

NGC 5643 ULX1 is a luminous ULX that is associated with the galaxy NGC 5643, which is at a distance at 16.9 Mpc. The source was initially identified using ROSAT/HRI data \citep{guainazzi04} at a luminosity of $\sim10^{40}$ erg s$^{-1}$. It was subsequently detected by \textit{Chandra} and \textit{XMM-Newton} with luminosities of  $\sim$1--4$\times 10^{40}$ erg s$^{-1}$, indicating the ULX exhibited long-term flux variability up to a factor of 2--3 (Figure~\ref{lc_long}).
We note that it is unlikely that this object is a background quasi-stellar object (QSO) as it does not have any detected optical counterpart in the DSS (down to $m_V\sim20$), in a short VLT/FORS1 exposure (down to a comparable magnitude). 
An optical counterpart would likely be detected in these observations if the source were a QSO with an observed X-ray flux of $\sim 1 \times 10^{-12}$ erg cm$^{-2}$ s$^{-1}$ (0.3--10 keV). We can place a lower-limit on the (0.3--3.5 keV) X-ray to optical flux ratio of $\sim$15, which is above the typical range of $f_X/f_V$ seen in most AGNs (0.1-10; e.g. \citealt{maccacaro88}). 
This is further supported by the lack of a detection of any potential counterpart to the ULX in infrared 2MASS and WISE images of the field.

NGC 5643 ULX1 was considered an intriguing source because, in low quality data, it appeared to show X-ray properties reminiscent of an IMBH accreting at sub-Eddington rates.  
Here we instead report that during a long {\it XMM-Newton} observation, the ULX had a spectrum which could be well modelled either in terms of a cold, optically thick Comptonizing corona or a very hot (2.5--3 keV), advection dominated disc. The high energy roll-over was further confirmed by a {\it NuSTAR} observation of the source \citep{annuar15}, indicating that the spectral shape of NGC 5643 ULX1 was more complex than the single powerlaw as found in low-quality data. {Furthermore, it is important to note that we observed a change in luminosity of a factor of 2--3 between the observations, without significant spectral changes.}

It is worth-noting that for most of the ULX population, the X-ray data are of poor quality and the paucity of the counting statistics can lead to an incorrect or uncertain interpretation of their nature. This was the case of NGC 5643 ULX1.
The properties showed by NGC 5643 ULX1 are not unusual for the ULXs.
Indeed, a number of other ULXs showed similar degeneracies between spectral shape, luminosity and short-term variability when comparing low and high quality data, and it has been proposed that these ULXs may be sMBHs or massive stellar BHs accreting at super-Eddington rates \citep[e.g.][]{stobbart06,roberts07,gladstone09,heil09,middleton11,sutton13,pintore14,middleton15}. 
If the power-law-dominated spectra of NGC 5643 ULX1 are indeed simply due to the poor data quality and the intrinsic spectra are actually similar to the highest quality observation, then the marginally different levels of short-term variability could be explained by the effects of an outflow ejected by the accretion disc at radii where it is locally super-Eddington \citep{poutanen07,ohsuga11,takeuchi14}. If the wind is turbulent, it could increase the short-term variability when our line of sight encounters turbulences in the wind \citep{takeuchi14,middleton15}. Therefore, if we believe to the very marginally different short-term variability observed at the same luminosity in two \textit{XMM-Newton} observations of this source, it could be explained as a substantial difference in homogeneity of the wind itself \citep{middleton15}. However, we note that if the outflow/wind extends above the whole disc, we may expect to observe its photosphere in the form of a cold thermal component.

If it is indeed a standard ULX, the spectral properties of NGC 5643 ULX1 may be associated with the {\it broadened disc} classification proposed by \citet{sutton13}.
However, we note that this classification may be ambiguous given the observed high X-ray luminosity and intermittent short-term variability, and this source could potentially be associated with the {\it hard ultraluminous} class. 
This could be the case if the source is seen close to face-on. We suggest that the X-ray spectrum could be dominated by the hard emission component, which is amplified by beaming in the funnel of the wind and scattered on the wind photosphere \citep[e.g.][]{middleton15}. 
In such conditions the soft spectral component may be suppressed, allowing the hard component to dominate, giving a harder spectrum (e.g. \citealt{sutton13}, \citealt{sutton13b}b). This scenario cannot be distinguished from an advection dominated disc, but we note that very marginal indications of a soft component are evident when the highest quality spectra are fitted with the multicolour disc plus Comptonisation model. 
Recently, a few other ULXs with luminosities $>10^{40}$ erg s$^{-1}$ have been reported to have single component, disc-like spectra \citep[Circinus ULX5, NGC 5907 ULX1; ][]{walton13b,sutton13b}, including one HLX candidate in the galaxy NGC 470, which appears to be disc-like when it is most luminous \citep{sutton12}. 
Similarly, several other well-studied lower luminosity ULXs have been reported to become spectrally harder with increasing luminosity (the NGC 1313 X-2 when it has a `very thick' spectra, \citealt{pintore12}; Holmberg IX X-1,  \citealt{vierdayanti10}, Luangtip et al. submitted). 
The feasibility of this scenario is seemingly confirmed by \citet{kawashima12}, 
who used Monte-Carlo simulations to simulate radiation spectra from supercritical BH accretion flows. They showed that a source accreting at the highest super-Eddington rates, where the wind envelope can cover the whole object and down-scatter the photons emitted in the hot inner regions, could appear with similar spectral properties to NGC 5643 ULX1 when viewed close to face-on.
This may naturally explain the (marginally) increased short-term variability in observation 1, whereas the potential origin of variability in a simple advection dominated disc is unclear. However, high levels of short-variability are generally seen in ULXs with a soft spectral shape, rather than hard sources such as NGC 5643 ULX1 \citep[e.g.][]{sutton13}. 
As such, we can speculate that in NGC 5643 ULX1 the wind does not completely cover the disc, but the hard component is beamed by the wind funnel so still dominates the observed spectrum. In this case, if clumps of wind material are blown in to the funnel and intersect our line of sight, they could feasibly produce the observed short-term variability. 

Based on the similarities between this source and NGC 1313 X-2, we tentatively estimated the mass of the compact object using the {\it hardness-intensity} diagrams from \citet{pintore14}. 
As NGC 5643 ULX1 displays similar spectral hardness to NGC 1313 X-2, we suggest that they may be accreting at similar fractions of Eddington. As NGC 5643 ULX1 is a factor of $\sim4$ more luminous than NGC 1313 X-2, for the same spectral hardness, we suggest that the BH in NGC 5643 ULX1 is also $\sim$4 times larger than that in NGC 1313 X-2. 
Indeed, at a luminosity of $4\times10^{40}$ erg s$^{-1}$, this ULX is consistent with containing a $\sim$30 M$_{\odot}$ BH which is emitting at $\sim$10 times the Eddington limit (cf. \citealt{poutanen07}).
This proposed scenario could be tested using further moderately long ($\sim$40 ks) combined \textit{XMM-Newton} and \textit{NuSTAR} observations, which could trace the evolution of the source. 

We must caution that the distance to the galaxy (hence the ULX distance as well) may be over estimated in this and previous work. 
\citet{sanders03} estimated a distance of 13.9 Mpc, based on the assumption of the cosmic attractor model \citep{mould00}. At this distance the luminosities of ULX1 would be reduced by $\sim$30\%. This would mean that the ULX is less extreme in luminosity than previously reported and possibly more similar to the bulk population of typical ULXs in the nearby Universe (e.g. \citealt{gladstone09,middleton11,vierdayanti10,pintore12,sutton13,pintore14,middleton15,walton13,bachetti13}). This would imply that our previous highly tentative mass estimate for the black hole relative to NGC 1313 X-2 should be lowered to a factor of $\sim$2--3.

\section{Conclusions}

We have analysed all of the available {\it XMM-Newton} observations of the SAB(rs)c galaxy NGC 5643, with the primary purpose of studying ULX1. 
This is an improvement over previous work, as for the first time we report results from a new observation with unprecedented data quality for this source.
The ULX is particularly luminous ($>10^{40}$ erg s$^{-1}$) and has an energy spectrum that can be reproduced by models of an advection dominated disc or a Comptonising corona, without showing significant spectral variability. Such a thermal-like spectral state is not seen in Galactic sub-Eddington accreting BH binaries, where soft spectra are generally well fitted by a standard Shakura \& Sunyaev accretion disc \citep{shakura73}. In addition, the short-term variability of this source appears to (very marginally) vary independently of the luminosity of the ULX. These results have allowed us to rule-out canonical sub-Eddington spectral states (i.e. sub-Eddington {\it hard} and {\it soft} states) which would be expected if the ULX contained an IMBH. Instead, we favour an interpretation of this source as BH of stellar origin accreting at super-Eddington rates. In particular, we suggest that the source may be seen almost face-on, with a strong outflow which beams and amplifies the hard emission produced in the inner disc. This is equivalent to the {\it hard/ultraluminous} state proposed by \citet{sutton13} for the highest luminosity ULXs. This scenario can not only produce the hard spectral shape seen in this ULX, but if the wind is turbulent, it also naturally explains the observed random short-term variability. Based on this interpretation, we tentatively estimate the mass of the BH in this system to be $\sim30$ M$_{\odot}$.

\section*{Acknowledgements} 
We thank the anonymous referee for his/her useful comments which helped us to improve the paper. We acknowledge financial support through INAF grant PRIN-2011-1 (Challenging ultraluminous X-ray sources: chasing their black holes and formation pathways) and INAF grant PRIN 2012-6. Based on observations obtained with \textit{XMM-Newton}, an \textit{ESA} science mission with instruments and contributions directly funded by \textit{ESA} Member States and \textit{NASA}. LZ acknowledges financial support from the ASI/INAF contract n. I/037/12/0. MJM appreciates support from ERC grant 340442. TPR was funded as part of the STFC consolidated grant awards ST/K00861/1 and ST/L00075X/1. ADS was supported by an appointment to the NASA Postdoctoral Program at Marshall Space Flight Center, administered by Universities Space Research Association through a contract with NASA.

\addcontentsline{toc}{section}{Bibliography}
\bibliographystyle{mn2e}
\bibliography{biblio}

\appendix
\section{Spectral Analysis of the AGN}
\label{AGN}

Besides the ULX, NGC 5643 hosts also an AGN. The AGN is known to be heavily absorbed and characterised by a large iron emission feature, probably associated with Compton reflection of hard photons (e.g. \citealt{guainazzi04, bianchi06, annuar15}). Other emission features have also been found in the {\it XMM-Newton} spectra of the AGN, and these have been associated with photoionised emission regions. \citet{annuar15} studied a combination of {\it NuSTAR} and {\it XMM-Newton} data with the aim of well characterising the spectral properties of the AGN. They found that different reflection/obscuration models can well describe the data, indicating that both Compton thick absorption by a toroidal absorber and reflection from distant, cold material are important in explaining the AGNs spectral properties. However, the properties of all of the components are still a matter of debate, and with the recent long {\it XMM-Newton} observation we have an unprecedented chance to study these features in high quality data at low energies ($<3$ keV).

We extracted the spectrum of the AGN in observation 3, choosing extraction regions of 20$''$ and 40$''$ for the AGN and the background, respectively. The spectra clearly show a prominent emission line 6.4 keV, plus many other emission features. Simple models, such as a blackbody plus a power-law or APEC plasma emission plus a power-law result in poor fits and broadband residuals. Instead we found that the \textit{XMM-Newton} continuum spectrum can be well fitted by the combination of a neutral absorbed {powerlaw}, a reflection component (modelled with {\sc reflionx}, \citealt[e.g.][]{ross07}) and a photoionised plasma ({\sc apec} model in {\sc xspec}). We fixed the chemical abundance of both the reflection and APEC plasma components to 1, and left the photon indices of the power-law and the ionising flux for the reflection component free to vary independently.
This three-component model provides an improvement in the fit to the data ($\chi^2/$dof=941/589, 
see Table~\ref{table_continuum}), with respect to the simpler models mentioned above, and the iron emission line is well modelled, allowing us to associate it with reflection. The neutral column density is $\sim0.4\times10^{22}$ cm$^{-2}$, which is intermediate between values reported for the previous {\it XMM-Newton} observations \citep[e.g.][]{guainazzi04,matt13}, but significantly lower than the value reported by \citet{annuar15}. The power-law has a photon index of $\sim5$ and can adequately reproduce the soft part of the spectrum with the additional contribution from the $\sim0.8$ keV APEC plasma. The very flat {powerlaw} photon index would give an unusually low Compton $y$ parameter; 
hence it may actually suggest that this component represents the tail of a thermal component which is outside of the \textit{XMM-Newton} bandpass. The reflection component mainly contributes to the high energy spectrum. From the inferred parameters, we can conclude that reflection occurs in a very-low ionisation medium ($\xi\sim20$ erg cm s$^{-1}$) illuminated by an ionising flux with a flat photon index. 
We note that the photon index is pegged at the lower limit allowed by the {\sc reflionx} model ($\Gamma=1.4$), although this may depend on the uncertainties in the illuminating emission, which is probably above the {\it XMM-Newton} bandpass. Hence we cannot exclude the possibility that the ionising flux spectrum is even flatter. 

\begin{table}
\footnotesize
\begin{center}
\caption{Best fitting model parameters for the AGN spectrum fitted with an absorbed three-component continuum {\sc apec+reflionx+powerlaw} model with six emission lines with moderate broadening (see text). Errors are at the 90$\%$ c.l. for each parameter.}
\label{table_continuum}
\begin{tabular}{lll}
\hline
\\
{\sc tbabs} 	    &$N_{\text{H}}$ (10$^{22}$ cm$^{-2}$)$^a$ &$0.372^{+0.02}_{-0.008}$ 		 \\
\\
{\sc apec} 	    &$kT$ (keV)$^b$			    & $0.79^{+0.04}_{-0.04}$	  \\
& Abundance$^c$ 							& =1 (fixed) \\
& Norm. ($10^{-5}$)$^d$ 						    & $3.8^{+0.1}_{-0.2}$	 \\
\\
{\sc powerlaw} &$\Gamma$$^e$	     			    & $4.7^{+0.4}_{-0.5}$ 		 \\
& Norm. $^f$ 						    		    & $2.1^{+0.4}_{-0.2}$ 			  \\
\\
{\sc reflionx} 	    & $Fe/solar$$^g$ 				   & =1 (fixed) 				  \\
 		   	    & $\Gamma$$^h$ 	   & 	$1.40_{*}^{+0.01}$ 	\\
 		   	    & Log $\xi$$^i$ 	   & 	$1.32_{-0.03}^{+0.05}$ 	\\
 		   	    & $Norm.$ ($10^{-6}$)$^j$ 	   	   & $12.2_{-0.4}^{+0.5}$ 	\\

\hline
		 	    &$\chi^2/dof$				   &686.61/571			\\
\end{tabular} 
\end{center}
\begin{flushleft} $^a$ Neutral column density; $^b$ photoionised plasma temperature; $^c$ chemical abundance of the plasma; $^d$ normalisation of the APEC plasma; $^e$ photon index;  $^f$ power-law normalisation; $^g$ ratio of the iron and hydrogen abundances; $^h$ photon index of the illuminating power-law spectrum; $^i$ logarithm of the ionisation parameter in units of erg cm s$^{-1}$; $^j$ normalisation of the {\sc reflionx} component.
\\
\end{flushleft}
\end{table}

\begin{figure}
\center
\includegraphics[width=6.5cm,height=8.5cm,angle=-90]{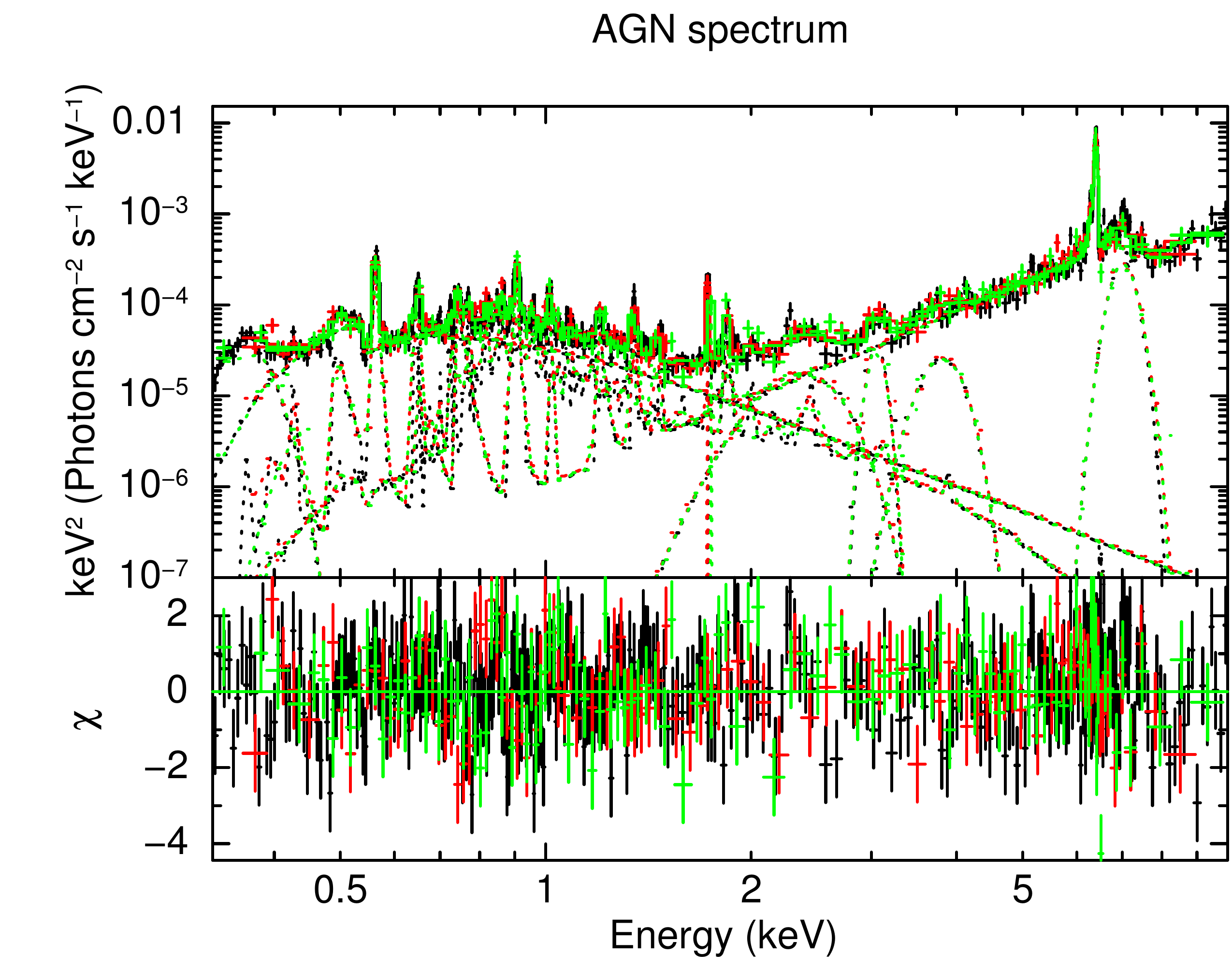}
\caption{EPIC-pn (black) and MOS (red and green) spectra of the AGN in NGC 5643, extracted from {\it XMM-Newton} observation 3. The spectra are unfolded from the detector response and fitted with an absorbed three-component continuum model ({\sc powerlaw+apec+reflonx}), with six emission lines.}
\label{AGN_spec}
\end{figure}

Emission lines are still evident as residuals in the spectrum. 
These include: moderately broad ($\sigma<70$ eV) emission lines at 0.32$\pm$0.3 keV, 1.73$\pm$0.1 keV and 3.05$\pm$0.5 keV (with equivalent widths, EWs, of $\sim$400 eV, $\sim$150 eV and $\sim$180 eV), associated with the neutral K$\alpha$ transition of C, Si and Ar, and  three broad ($\sigma\sim$250 eV) emission features 2.29$\pm$0.8 keV, 3.8$\pm$0.1 keV and 7.00$\pm$0.7 keV (with EWs of 520 eV, 220 eV and 500 eV respectively), associated with neutral K$\alpha$ transitions of N and Ca, and the highly ionised Fe XXVI K$\alpha$ line respectively (see Figure~\ref{AGN_spec}). 
The best fitting model that includes the emission lines results in a fit statistic of $\chi/dof=687/571$, which is formally acceptable (see table~\ref{table_continuum}).
We suggest that the C, Si, Ar, Ca and N lines are related to the reflection component which produces the intense 6.4 keV Fe emission line, whilst the Fe XXVI emission line may be generated in a very ionised region close to the central supermassive BH.
We infer a 0.3--10 keV unabsorbed luminosity of $3\times10^{41}$ erg s$^{-1}$, thus confirming that the source is a low luminosity AGN (e.g. \citealt{guainazzi04,matt13,annuar15}).

The spectral analysis shows that the AGN has a complex spectral shape that can be well modelled in the soft band by a photoionised plasma and a steep power-law with photon index of 5. The plasma emission has a temperature of $\sim0.8$ keV and an unabsorbed 0.3--10 keV luminosity of $\sim$3$\times10^{39}$ erg s$^{-1}$, and may originate from gas heated by shocks or winds in regions of intense star formation \citep{guainazzi04}. 
These two components could be associated with an optically thin medium located close to the X-ray source, 
in which the tail of a cold thermal component from outside of the {\it XMM-Newton} bandpass is reprocessed, mainly through photoelectric absorption.
The hard energy band is dominated by a reflection component that appears to be produced by the scattering of hard photons (with a spectral index of 1.4) in a low-ionised medium (almost neutral). Some broad emission features are also observed, which we identify as K$\alpha$ neutral transitions of C, Si, Ar, Ca and N. These emission features are probably associated with the same reflection processes which dominate the high energy continuum emission. The broadening (up to 0.3 keV) could be caused by either scatterings in a high-velocity medium, thermal Compton scattering or by relativistic effects. 
Our results are consistent with the broadband spectral study presented by \citet{annuar15}, who found that a combination of highly Compton thick absorption ($>10^{24}$ cm$^{-2}$) and cold reflection are at work in this source. The 2.0--10 keV unabsorbed luminosity found in observation 3 ($\sim2.5\times10^{40}$ erg s$^{-1}$) is clearly lower than the $0.8-1.7\times10^{42}$ erg s$^{-1}$ reported by \citet{annuar15}, and we relate this discrepancy to both the significantly lower absorption column found in this work and to possible real slight variation in the X-ray luminosity. \citet{annuar15} showed that a model which take into account only reflection converges towards low values of the column density. Instead, a combination of toroidal absorption and reflection may result in more reliable estimates of the column density.

\end{document}